\documentclass[a4paper,11pt]{article}
\usepackage{jheppub}
\usepackage{epsfig,epsf}
\usepackage{mathrsfs}
\usepackage{graphicx}
\usepackage{latexsym}
\usepackage{amsmath}
\usepackage{amssymb}
\usepackage{textcomp}
\usepackage{amsbsy}
 \usepackage[normalem]{ulem}
 
\renewcommand{\baselinestretch}{1.45}

\def\beq{\begin{equation}}
\def\eeq{\end{equation}}
\def\barr{\begin{array}}
\def\earr{\end{array}}
\def\dis{\displaystyle}
\def\tev{\, {\rm TeV}}

\def\lapp{\mathrel{\rlap{\raise.5ex\hbox{$<$}}
                    {\lower.5ex\hbox{$\sim$}}}}
\def\gapp{\mathrel{\rlap{\raise.5ex\hbox{$>$}}
                    {\lower.5ex\hbox{$\sim$}}}}
\def\sup{{\cal S}_{\rm up}}
\def\sdn{{\cal S}_{\rm dn}}
\def\ft{\widetilde f}
\def\gym{g_{\mbox{\lower.25ex\hbox{\tiny YM}}}}

\title{Bulk gauge and matter fields in nested warping: \\I. the formalism}

\author[1]{Mathew Thomas Arun,\note{Corresponding author.}}
\author{Debajyoti Choudhury}
\emailAdd{thomas.mathewarun@gmail.com}

\emailAdd{debajyoti.choudhury@gmail.com}
\affiliation[a]{Department of Physics and Astrophysics, University of Delhi, 
Delhi 110007, India.}
\abstract
{The lack of evidence for a TeV-mass graviton has been construed
  as constricting the Randall-Sundrum model. However, a doubly-warped
  generalization naturally avoids such restrictions. We develop, here,
  the formalism for extension of the Standard Model gauge bosons and
  fermions into such a six-dimensional bulk. Apart from ameliorating
  the usual problems such as flavour-changing neutral currents, this
  model admits two very distinct phases, with their own unique
  phenomenologies.
}
\begin{document}
\maketitle
\flushbottom


\section{Introduction}
   \label{sect:intro}
The spectacularly successful Standard Model (SM) has received a recent
fillip with the discovery of the long-sought for Higgs
boson~\cite{atlas-higgs, cms-higgs}. Yet, certain questions remain
unanswered. These pertain to the existence of Dark Matter, the origin
of the baryon asymmetry in the Universe, the existence of multiple
generations of fermions, the hierarchy in fermion masses and mixing,
and, last but not the least, the stability of the Higgs sector under
quantum corrections. Over the years, several attempts have been made
to answer these questions, albeit only with partial success. One such
stream of thought envisages a world in more than three space
dimensions as a possible panacea to some of the ills of the SM, and,
in this paper, we concentrate on this possibility.

While such theories were first proposed nearly a century
ago~\cite{nord,kaluza,klein} in the quest to unify electromagnetism
and gravity, the early efforts were quickly shown to lead to a dead
end and were abandoned. The situation changed with the introduction of
String Theory as a quantum theory of gravity as well as a possible
ultraviolet completion of the SM. With the theory defined, of
necessity, in at least ten dimensions, compactification of the extra
dimensions is paramount before it can be deemed a description of the observed
world. With the compactification scale being close to the Planck scale
in most early constructions, the new dimensions, understandably played
virtually no role in low-energy physics.  However, warped
compactification~\cite{RS1,RS2}, wherein the SM fields were confined
to the usual $(3 + 1)$ dimensions with only gravity being allowed to
propagate in the (five-dimensional) bulk, led not only to a
``resolution'' of the hierarchy problem, but also to interesting
consequences at colliders owing to distinct ${\cal O}(\tev)$
resonances in the form of
Kaluza-Klein (KK) excitations of the
graviton.

The last feature moved both the ATLAS~\cite{Aad:2012cy} and
CMS~\cite{Khachatryan:2014gha} experiments to investigate the
existence of such Randall-Sundrum (RS) graviton resonances, especially
through the dilepton and diphoton decay channels. In particular, the
ATLAS experiment ruled out graviton masses below $1.03~(2.23) \tev$ at
95\% C.L. with the lower bound being dependent on the ratio of the
five-dimensional curvature and the fundamental mass scale. This ratio is constrained, on the upper
side, by the applicability of a semiclassical treatment (only recourse
available in the absence of a full quantum theory of gravity) and, on
the lower, by the undesirability of fine-tuning. For reasonable values
of this ratio, the mass of the first graviton excitation should,
preferably, be a few times that of the Higgs boson; and certainly
no higher than a few TeVs. Thus, the continued absence of any such
resonance at the TeV scale begins to call into question the validity
of this scenario as a cure for the Higgs mass stabilization problem.
However, it should be realized that the RS
model is only the simplest of possible warped world
scenarios. In particular, there is no reason that there should be only
one such extra dimension~\cite{Shaposhnikov,Nelson,Cohen,Giovannini,
Kanti,ChenF,Choudhury:2006nj}. For one, in a scenario with
double (or more) warping~\cite{Choudhury:2006nj}, it was shown
recently that the aforementioned ATLAS bounds are naturally
evaded~\cite{Arun:2014dga}. This motivates us to study the features
of six-dimensional theories.

Although the mechanism for the formation of branes and the
localization of fermions thereon is well understood, it is interesting
to consider allowing them to propagate in the full six dimensions. The
corresponding flat space theories have several interesting
consequences.  For example, the analogue of Witten anomaly
cancellation leads to a prediction of the number of chiral
generations~\cite{Dobrescu:2001ae}, while suppressing the proton decay
rate to below the current constraints~\cite{Appelquist:2001mj}.
Furthermore, some of these constructions~\cite{Rubakov:1983bz}
naturally lead to a small cosmological constant. The rich collider
phenomenology~\cite{Burdman:2006gy,Dobrescu:2007xf,Freitas:2007rh,Choudhury:2011jk},
apart from the existence of a viable cold dark matter candidate
\cite{Dobrescu:2007ec,Cacciapaglia:2009pa} renders these scenarios
phenomenologically attractive.  On the other hand, with the KK excitations for each
species now expanding to a ``tower of towers'', the quantum
corrections to the SM amplitudes ---most importantly to the
electroweak precision variables--- are potentially large, calling into question the
consistency with low-energy phenomenology.  However, as
ref.~\cite{chang} demonstrated for a five-dimensional theory, it is
possible to suppress the coupling between the
zero modes of the SM fields
and the KK towers. This is of particular importance in the context of
the aforementioned quantum corrections. Thus, it is of interest to
investigate whether considering a warped space would allow us to
preserve some of the advantages of going into six dimensions while
simultaneously protecting us from the pitfalls. This paper is the
first step towards this goal, and we set up the entire formulation
here and comment
 on some of the consequences. While
ref.\cite{Das:2011fb} did consider bulk SM fields in such a geometry,
the analysis therein had taken recourse to an approximation of the
metric, thereby leading to a significant simplification of the
equations of motion. However, as was demonstrated in
ref.\cite{Arun:2014dga}, the said approximation, apart from being
untenable close to the brane we live on, led to a drastic change in
the form of the graviton wavefunctions (and, hence, their couplings).
As we shall show, much the same happens for bulk gauge bosons as well,
leading to very interesting phenomenological consequences. The
detailed phenomenology would be presented in subsequent
papers.\footnote{It should be pointed out that six dimensional warped
models with spherical compactifcations \cite{Gogberashvili:2003xa,
Gogberashvili:2003ys}, do try to explain the number of fermion
families \cite{Gogberashvili:2007gg}.  However, with these models
having only a single warping, the aforementioned constraints on the RS
scenario continue to hold, albeit with some modifications.}

The rest of the article is constructed as follows. To begin with, we
present a very brief review of the doubly warped space. Sections
\ref{sec:fermions}\&\ref{sec:gauge} discuss, respectively, the bulk
fermions and gauge bosons in this theory, without taking into
consideration the spontaneous breaking of the gauge
symmetry which, in turn, is discussed in Sec.\ref{sec:higgs}.
The interactions are delineated in Sec.\ref{sec:interactions} and the Feynman 
rules listed. Finally, we conclude in Sec.\ref{sec:summary}.

\section{Brief review: nested warping in six dimensions}    
We consider a compactified six-dimensional space-time with successive
warpings and $Z_2$ orbifolding in each of the two extra dimensions,
viz.  $M^{1,5}\rightarrow[M^{1,3}\times S^1/Z_2]\times
S^1/Z_2$. Dual requirements of nested warping along with 
a manifestly exhibited four-dimensional ($x^\mu)$
Lorentz symmetry restricts the line element to the form~\cite{Choudhury:2006nj}
\begin{equation}
ds^2= b^2(x_5) \, [a^2(x_4)\eta_{\mu\nu}dx^{\mu}dx^{\nu}+R_y^2dx_4^2]
     +r_z^2dx_5^2 \ ,
\label{metric}
\end{equation}  
where the compact directions are represented by the dimensionless
coordinates $ x_{4,5}\in [0,\pi]$ with $R_y$ and $r_z$ being the
corresponding moduli. It is interesting to examine
  the rationale for the two orbifoldings.  A nontrivial $a(x_4)$, when
  accompanied by compactification, demands (as in the RS case), the
  orbifolding in the $x_4$-direction. Furthermore, it necessitates
the presence of localized energy densities at the orbifold fixed
points, and in the present case, these appear in the form of tensions
associated with the two end-of-the-world 4-branes at $x_4 = 0,
\pi$. Similarly, 
even without any orbifolding in the $x_5$-direction, a nontrivial
$b(x_5)$ for a compactified $x_5$ automatically requires that a
4-brane should exist at $x_5 = \pi$, whereas none needs to exist at
$x_5 = 0$.
The situation changes though if one wishes to introduce such a
  brane. While the latter
  could exist even in the absence of such an
  orbifolding, it would be free to traverse in the $x_5$-direction
  in the absence of a constraining potential. Thus, if such a brane is
  to be introduced, it is easiest to do so if the second $S^1$ is
  orbifolded too.

The total bulk-brane action for the six
dimensional space time is, then, given by
\begin{equation}
\barr{rcl}
{\cal{S}}&=& \dis {\cal{S}}_6+{\cal{S}}_5 \\[1.5ex]
{\cal{S}}_6&=& \dis \int d^4x \, dx_4 \, dx_5 \, \sqrt{-g_6} \, 
   (M_{6}^4R_6-\Lambda)\\[1.5ex]
{\cal{S}}_5&=& \dis \int d^4x \, dx_4 \, dx_5 \, \sqrt{-g_5}\, 
      [V_1(x_5) \, \delta(x_4)+V_2(x_5) \, \delta(x_4-\pi)]\\
&+& \dis \int d^4x \, dx_4 \, dx_5 \, \sqrt{-\tilde g_5} \, 
     [V_3(x_4) \, \delta(x_5)+V_4(x_4) \, \delta(x_5-\pi)] \ ,
\earr
    \label{6d_action}
\end{equation}
where $\Lambda$ is the (six dimensional) bulk cosmological constant and 
$M_6$ is the
fundamental scale (quantum gravity scale) in six dimensions. The
five-dimensional metrics in ${\cal S}_5$ are those induced on the
appropriate 4-branes which lend a rectangular box shape to the
space.  

If the bulk cosmological constant $\Lambda$ is negative, the solutions
for the 6-dimensional Einstein field equations are given
by~\cite{Choudhury:2006nj}
\begin{equation}
\barr{rclcrclcrcl}
a(x_4)& = & e^{-c|x_4|} &\hspace{2cm}& c & = & \dis \frac{R_y k}{r_z\cosh{k\pi}} & \equiv & \dis \frac{\alpha \, k}{\cosh(k \pi)}
\\[1ex]
b(x_5)& = & \dis \frac{\cosh{(k x_5)}}{\cosh{(k\pi)}} &\hspace{2cm}& 
k& = & \dis r_z\sqrt{\frac{-\Lambda}{10 M_6^4}} 
& \equiv  & \dis 
\epsilon \, r_z \,M_6 \ ,
\earr
                \label{RS6_eqns}
\end{equation}
where we have introduced the dimensionless constants $\alpha$ and 
$\epsilon$ for future reference. Clearly, the validity of the semiclassical 
treatment (to the extent of neglecting quantum corrections to the bulk gravity 
action) requires the bulk curvature to be significantly smaller than the 
fundamental scale $M_6$ and it has been argued in the literature to imply 
that $\epsilon \lapp 0.1$.
Similarly, the ratio of the two moduli 
should not be too large so as to not reintroduce a large 
hierarchy.

As in the RS scenario, the brane tensions in eqn.(\ref{6d_action}) 
are specified by the junction conditions.
The smoothness of the warp factor
at $x_5 = 0$ obviates the necessity for a $V_3(x_4)$, while the fixed
point at $x_5 = \pi$ requires a negative tension, viz.
\begin{equation}
\label{zten}
V_3(x_4)=0,\hspace{1cm} V_4(x_4)=\frac{- 8M_6^4k}{r_z} \,\tanh{(k \pi)} \ .
\end{equation}
In contrast, the two 4-branes sitting at 
$x_4 = 0$ and $x_4 = \pi$ require equal and
opposite energy densities, just as in the RS case. However, 
the $x_5$-warping dictates 
that, rather than being constants, these energy densities must 
be $x_5$-dependent, viz.
\begin{equation}
\label{yten}
V_1(x_5)=-V_2(x_5)=8M_6^2\sqrt{\frac{-\Lambda}{10}} \, {\rm sech}(k x_5) \ .
\end{equation}
Such tensions could originate from different kinds of
physics~\cite{Choudhury:2006nj}. The most simple would be a scalar
field with a non trivial potential, wherein the solution of
eqn.(\ref{RS6_eqns}), along with the expressions for $V_{1,2}(x_5)$, is
nothing but a self consistent solution for the gravity--scalar
system. While a wide variety of potentials can have such a kink-like
solution~\cite{Choudhury:2006nj}, it is intriguing to note that 
even a simple quartic form can do the job.
(In Sec.\ref{sec:higgs}, we shall
encounter a variant of this mechanism.) And while other scenarios,
such as a Born-Infeld action, can also lead to a similar effective potential,
we shall not explore those here.

The (derived) 4-dimensional Planck scale can be related 
to the fundamental scale $M_6$ through 
\begin{equation}
   M_{\rm Planck}^2 \sim \frac{M_6^4 \, r_z \, R_y}{2 \, c \, k}  \,
              \left(1 - e^{-2 \, c \, \pi} \right) \;
              \left[ \frac{\tanh k \pi}{\cosh^2 k \pi} 
                   + \frac{\tanh^3 k \pi}{3}  \right] \ .
    \label{Planck_mass}
\end{equation}
The point in this 2-dimensional ($x_4$--$x_5$) plane 
associated with the lowest energy scale is given by 
$x_4 = \pi, x_5 = 0$. Assuming us to be located at this juncture 
immediately gives the required hierarchy factor ({\em i.e.}, the mass
rescaling due to warping) to be
\begin{equation}
\label{hierarchy}
w =  \frac{e^{-c \pi}}{\cosh{k\pi}} \ .
\end{equation}

\begin{figure}[!h]
{
\vspace*{-30pt}
\centerline{\epsfxsize=8cm\epsfbox{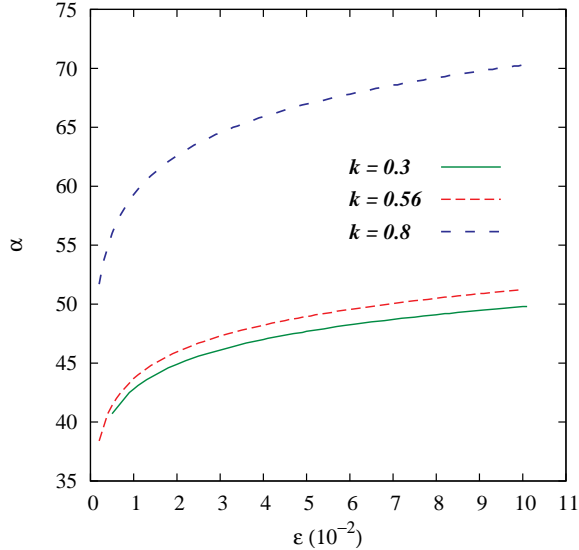}}
\vspace*{-60pt}
}
\caption{\em Contour plots in the $(\epsilon, \alpha)$ plane
  for fixed values of $k$ with $R_y$ set to satisfy the hierarchy eqn.(~\ref{hierarchy}).}
\label{fig:eps_alpha}
\end{figure}

For the large $w$ that we need, this equation,
along with the relation between $c$ and $k$ (eqn.\ref{RS6_eqns})
demands that, unless there is a very large hierarchy between the
moduli, the warping is substantial in only one of the two directions,
and rather sub-dominant in the other.  In other words, we can have
either ($i$) a large ($\sim 10$) value for $k$ accompanied by an
infinitesimally small $c$ or ($ii$) a large ($\sim 10$) value for $c$
with a moderately small ($\lapp 1.0$) $k$. The relationship between the 
parameters of the theory are displayed in Fig.\ref{fig:eps_alpha}.

In summary, we are dealing with a brane world which is doubly warped,
with the warping being large along one direction and small in the
other.  The very structure of the theory typically requires a small
hierarchy between the two moduli, both of which remain comparable to
the fundamental length scale in the theory. It should be realized that 
the two branches, namely $(i)$ a large $k$ and a small $c$, or 
$(ii)$ a large $c$ and a small $k$, are fundamentally different.

\section{The fermions}
    \label{sec:fermions}
As is well-known, fermions, of necessity, are defined as 
representations of the Poincare algebra, as applicable to the
tangent space. Hence, we begin by briefly reviewing the construction
in flat space before embarking on the more germane issue of the warped 
space.
\subsection{Flat six-dimensions}
In the current case, the spin-$1/2$ representation 
is defined by
six $8 \times 8$ matrices $\Gamma^a$, satisfying a Clifford algebra 
\[
\left\{\Gamma_a, \Gamma_b \right\} = 2 \, \eta_{ab}
\]
where $\eta_{ab}= {\rm diag}(-1,+1,+1,+1,+1,+1)$. We choose to work with
a particular representation of the algebra defined by 
\beq
\barr{rcl c rcl}
\Gamma_{\hat \mu} & = & \gamma_{\hat \mu} \otimes \sigma_{3}
& \qquad & \Gamma_{4} & = &  \, 1 \otimes \sigma_{1}\\
\Gamma_{5} & = &  \, 1 \otimes \sigma_{2} 
& & 
\Gamma_{7} & = & \gamma_{5} \otimes  \sigma_{3} \ , 
\earr
\eeq
where $\gamma_{\hat \mu}$ (with $\hat \mu$ denoting the subspace of
the flat space) are the four-dimensional Dirac matrices (in the Weyl
representation) and $\gamma_5$ ($\Gamma_7$) is the parity operator in
four (six) dimensions\footnote{Note that this implies that
    $\Gamma_0$ is antihermitian, while the other $\Gamma_M$ are
    hermitian.}. The direct product is defined in a trivial sense
and, for example,
\[
\gamma_{\hat \mu} \otimes \sigma_{3} = 
    \left( 
    \barr{rcl}
         \gamma_{\hat \mu} && 0 \cr 
         0 && - \gamma_{\hat \mu}  
    \earr 
    \right) \ .
\]
It is straightforward to construct the Dirac spinor 
$\Psi(x_{\hat \mu},x_4,x_5)$ satisfying the flat space Dirac equation
\begin{equation}
\label{eq:Dirac}
\left[\Gamma_{\hat \mu} \partial^{\hat\mu}+\Gamma_{4}\partial_4
  +\Gamma_{5} \partial_5 - m\right] \Psi(x_{\hat \mu},x_4,x_5) = 0 \ .
\end{equation}
The representation of
the Lorentz generators given by
$\Sigma_{ab}=\frac{i}{2}[\Gamma_{a},\Gamma_{b}]$, viz.
\beq
\barr{rclcl r rclcl}
\Sigma_{\hat\mu\hat\nu}& = & \dis \frac{i}{2} [\Gamma_{\hat\mu},\Gamma_{\hat\nu}] & = & \dis 
    S_{\hat\mu\hat\nu} \otimes \sigma_{0} 
& \qquad & 
\Sigma_{\hat\mu4} & = & \dis \frac{i}{2}[\Gamma_{\hat\mu},\Gamma_{4}]
& = & - \gamma_{\hat\mu} \otimes \sigma_{2} \\[2ex]
\Sigma_{\hat\mu5} & = & \dis \frac{i}{2}[\Gamma_{\hat\mu},\Gamma_{5}]
& = & \gamma_{\hat\mu} \otimes \sigma_{1}
& & 
\Sigma_{45}& = & \dis 
\frac{i}{2}[\Gamma_{4},\Gamma_{5}] & = & \dis  1 \otimes \sigma_{3}
\earr
\eeq
is, of course, reducible, as is the case for all even dimensions. In
other words, this space admits chiral representations $\Psi_\pm$ and
thus, we may directly export the SM quantum number assignments. This
is quite unlike the five-dimensional case. Of course, on
compactification, each such six-dimensional chiral representation
would, in general, yield low-lying states carrying either value for the
four-dimensional chirality unless boundary conditions (such as those
pertaining to the orbifold fixed points) prevent this. Before we
consider such details, we need to set up the Dirac equation in the
warped six-dimensional space which we do next.


\subsection{Fermions in the warped space}

To define these, we need to consider the sechsbeins (namely, the
transformations to the tangent space) $e_M^a$ which satisfy the
conditions
\[
e_M^a \, e_N^b \, g^{MN} = \eta_{a b} \ , \qquad
e_M^a \, e_N^b \, \eta^{ab} = g_{MN}  \ ,
\]
leading to
\beq
e_\mu^a = a(x_4) \, b(x_5) \, \delta_\mu^a \ , \quad 
e_4^a = R_y \, b(x_5) \, \delta_4^a \ , \quad
e_5^a = r_z \, \delta_4^a \ .
\eeq
Denoting the inverse sechsbeins by $E^M_a$, we define the spin
connections $\omega_{bcM}$ through the covariant derivatives of
$E^M_a$, viz.,
\beq
\omega_{bcM} \equiv E_b^N \, (g_{PN} \, E_c^P)_{; M} 
      = g_{RN} \, E_b^N \, (\partial_M \, E_c^R + \Gamma^R_{MT} \, E_c^T) \ .
\eeq
For the metric of eqn.(\ref{metric}), the only nontrivial components 
of the spin connections are given by
\beq
\omega_{bc4} = \frac{R_y}{r_z} \, \dot b \, \delta^5_{[b} \, \delta^4_{c]} \ ,
\quad
\omega_{bc\mu} = \eta_{\mu\nu} \, \left( \frac{a'}{R_y} \, 
                 \delta^\nu_{[b} \, \delta^4_{c]} 
               + \frac{a \, \dot b}{r_z} \, 
                 \delta^\nu_{[b} \, \delta^5_{c]}  \right) \ ,
\eeq
where primes (dots) denote derivatives with respect to $x_4 \, (x_5)$.
The Dirac Lagrangian in the warped geometry is, then, given by 
\begin{equation}
{\cal L}_{\rm Dirac} = i \, \bar{\Psi}_{+} \, \Gamma^{a} \, E^{M}_{a} \, 
    \left( \partial_M + w_{M}^{bc}[\Gamma_b,\Gamma_c] \right) \, \Psi_{+}
\end{equation}
for the positive chirality field $\Psi_{+}$ and, analogously, 
for $\Psi_{-}$ as well. The corresponding equation of motion is 
\[
\Gamma^{a}E_{a}^{M}D_{M} \Psi_{+}\, = \, 
\left[\left(\frac{\Gamma^{\mu}}{ab}\partial_{\mu} + \frac{\Gamma^{4}}{R_y b}
      \partial_4
     +\frac{\Gamma^{5}}{r_z}\partial_5\right)
     +\frac{1}{2}\left(4 \Gamma^{4}\frac{a'}{ab R_y}
                  +5 \Gamma^{5}\frac{\dot{b}}{b r_z}\right)  \right] \Psi_{+} = 0 \ .
\]
Anticipating Kaluza-Klein reduction, we write the 
positive chirality Weyl spinor as 
\beq
\Psi_{+} = \frac{1}{\sqrt{R_y r_z}} \, \sum_{n,p}
     \left[{\cal F}^{n,p}_{l}(x_4,x_5) \, \psi^{n,p}_{l}(x_{\mu}) \otimes \ \sup
       + {\cal F}^{n,p}_{r}(x_4,x_5) \, \psi^{n,p}_{r}(x_{\mu})\otimes \sdn \right] \ ,
\eeq
with
\beq
\sup \equiv (1 \quad 0)^T \ , \quad
\sdn \equiv (0 \quad 1)^T \ . 
\eeq
Here, ${\cal F}^{n,p}_{r}(x_4,x_5)$ encapsulate the wavefunction
dependences on the extra dimensions, with the subscripts ($l,r$)
referring to the (four-dimensional) chirality of the putative
four-dimensional fields $\psi^{n,p}_{l,r}$ whereas the factor
$\sqrt{R_y \, r_z}$ (ensuring the correct mass dimension) would have arisen if 
the compactified directions were flat instead.  The Dirac
equation then reduces to
\beq
\barr{rcl}
0 & = & \dis 
\left[ \left(\frac{\gamma^{\mu}}{a}\partial_{\mu} \psi_{l}^{n,p}\right) \,
          {\cal F}^{n,p}_{l}
+ \psi_{r}^{n,p} \, 
       \left\{\frac{1}{R_y }\left(\partial_4 + 2 \frac{a'}{a}\right)
            -i \, \frac{b}{r_z} \left(\partial_5 + \frac{5 \,\dot{b}}{2 \,b}
                            \right) \right\} \, {\cal F}^{n,p}_{r} 
   \right] \otimes \sup
\\[2ex]
&+ & \dis
\left[-\left(\frac{\gamma^{\mu}}{a}\partial_{\mu} \, \psi_{r}^{n,p}
         \right) \, {\cal F}^{n,p}_{r} \, 
+ \psi_{l}^{n,p}
 \, \left\{ \frac{1}{R_y }\left(\partial_4 + 2 \frac{a'}{a}\right)
            +i \,\frac{b}{r_z} \, \left(\partial_5
                       + \frac{5 \, \dot{b}}{2 \, b} \right) \right\} \,
         {\cal F}^{n,p}_{l} 
\right] \otimes \sdn  \ .
\earr
\eeq

Expectedly, the two Weyl fields $\psi^{n,p}_{l/r}$ (at each level)
combine to give a Dirac fermion, and this results in
\begin{equation}
\barr{rcl}
 \gamma^{\mu}\partial_{\mu}\psi^{n,p}_{l/r} & = & \dis M_{n,p} \psi^{n,p}_{r/l}
\\
0 & = & \dis
\frac{M_{n,p}}{a} {\cal F}^{n,p}_l 
+ \frac{1}{R_y}\left(\partial_4 + 2 \frac{a'}{a}\right) {\cal F}^{n,p}_r 
-i \frac{b}{r_z}\left(\partial_5 + \frac{5}{2} \frac{\dot{b}}{b} \right)
{\cal F}^{n,p}_r 
\\
0 & = & \dis 
\frac{-M_{n,p}}{a}{\cal F}^{n,p}_r + 
\frac{1}{R_y}\left(\partial_4 + 2 \frac{a'}{a}\right) {\cal F}^{n,p}_l 
+i \frac{b}{r_z}\left(\partial_5 + \frac{5}{2} \frac{\dot{b}}{b}\right)
{\cal F}^{n,p}_l  \ .
\earr
\label{a1}
\end{equation}
Effecting a separation of variables, we write
\begin{equation}
{\cal F}_{l/r}^{n,p}(x_4,x_5) 
= \left[a(x_4)\right]^{-2} \, \left[b(x_5)\right]^{-5/2}
\ft^{n,p}_{l/r}(x_4)f_{l/r}^{p}(x_5)
\end{equation}
as this particular parametrization not 
only removes the spin connection terms from the equations of motion,
but also effectively isolates the 
derivative discontinuities in the wavefunctions at the 
boundaries. This leads to
\beq
\barr{rcl}
\dis \left(\frac{1}{R_y}\partial_4 -i \frac{b}{r_z}\partial_5\right)\ft^{n,p}_r(x_4)f_{r}^{p}(x_5)
 + \frac{M_{n,p}}{a} \ft^{n,p}_l(x_4)f_{l}^{p}(x_5) =0
\\[2ex]
\dis 
\left(\frac{1}{R_y}\partial_4 +i \frac{b}{r_z}\partial_5\right) \ft^{n,p}_l(x_4)f_{l}^{p}(x_5)-\frac{M_{n,p}}{a} \ft^{n,p}_r(x_4)f_{r}^{p}(x_5) =0 \ .
\earr
\eeq
Clearly, $f_{r/l}^p(x_5) = 1$ and $\ft_{r/l}^{n,p}(x_4) = 1$ 
satisfy the above for $M_{n,p} = 0$ and these, if permitted by 
the boundary conditions, would denote the 
ground state.

For nonzero $M_{n,p}$, these coupled equations can be diagonalized in a fashion analogous
to that for the flat space case, albeit at the cost of introducing 
slightly more complicated operators, viz.
\begin{equation}
\label{a11}
\barr{rcl}
0 & = & \dis (aD^- a D^+ + M_{n,p}^2) \, \ft^{n,p}_r(x_4)f_{r}^{p}(x_5) 
\\
0 & = & \dis (a D^+ a D^- + M_{n,p}^2) \ft^{n,p}_l(x_4)f_{r}^{p}(x_5) 
\\
D^\pm
& \equiv & \dis \frac{1}{R_y} \partial_4 \mp i \frac{b}{r_z}\partial_5  \ .
\earr
\end{equation}
On separating, these yield
\beq
\barr{rcl}
\label{fermioneqmotion}
0 & = & \dis 
a(x_4) \, \partial_4 \left[a(x_4) \, \partial_4 \ft^{n,p}_{l/r}(x_4)
\right]
+ R_y^2 \, \left[M_{np}^2 -m_{p}^2 a^2(x_4)\right]\ft^{n,p}_{l/r}(x_4)  
\\[1ex]
0 & = & \dis b(x_5)\partial_5\left(b(x_5) \partial_5 + i c \, 
{\rm sgn}(x_4) \,
  \frac{r_z}{R_y} \right)f_{r}^{p}(x_5)+m_p^2r_z^2f_{r}^{p}(x_5)  
\\[1ex]
0 & = & \dis b(x_5)\partial_5\left(b(x_5) \partial_5 - i c \,
{\rm sgn}(x_4) \,
\frac{r_z}{R_y} \right)f_{l}^{p}(x_5)+m_p^2r_z^2f_{l}^{p}(x_5) \ .
\earr
\eeq
Clearly, the $x_5$-equations can be factorized and their solutions,
in the bulk, would satisfy\footnote{The ostensible
    derivative discontinuities which would, putatively, have exchanged
    $f_l$ and $f_r$ at the boundary, is actually of no consequence at
    all as the physical range corresponds to $x_4 \geq 0$.}

\[
b(x_5) \, \partial_5 f_{l/r} = 
\left[ i \, \frac{c \, r_z}{2 \, R_y} \, \kappa_{l/r} \right]\, f_{l/r}
\]
where the exponent has been factorized for future 
convenience. The constants $\kappa_{l/r}$ are solutions of quadratic 
equations, and are given by
\beq
\barr{rcl}
\kappa_r^{\pm} & = & \dis 
-1 \pm \sqrt{1+4\frac{m_p^2R_y^2}{c^2}} 
\\[2ex]
\kappa_l^{\pm} & = & \dis 
1 \pm \sqrt{1+4\frac{m_p^2R_y^2}{c^2}}
\earr
\eeq
This leads to
\beq
\barr{rcl}
\label{flfr}
f_l(x_5) & = & 
  d_l^+
   \, \exp\left[i \kappa_{l}^{+} \Theta_k(x_5)\right] 
+   d_l^-
\, \exp\left[i \kappa_{l}^{-} \Theta_k(x_5)\right] 
\\[1ex]
f_r(x_5) & = & 
  d_r^+
\, \exp\left[i \kappa_{r}^{+} \Theta_k(x_5)\right] 
+   d_r^-
 \, \exp\left[i \kappa_{r}^{-} \Theta_k(x_5)\right] 

\\[1ex]
\Theta_k(x_5) & \equiv & \dis \tan^{-1} \, \left(\tanh \frac{k x_5}{2}\right) \ .
\earr
\eeq
To determine the constants $d_{l/r}^\pm$, we need to impose the 
boundary conditions, which, for phenomenological reasons, must be 
different for each chiral projection. To be specific, for the $SU(2)_L$-doublet
fields, we impose Neumann conditions for $f_l$ and Dirichlet for $f_R$. This, later on, would ensure that the zero-mode four-dimensional fermion
would be a left-handed field\footnote{For the  $SU(2)_L$-singlets, the condition would be opposite resulting in only the right-handed component having a zero-mode.}. In other words, we demand
$\partial_5f_l|_{x_5=0,\pi} = 0 $ and $f_r|_{x_5=0,\pi} = 0$. Thus,
\[
\barr{rclcrcl}
\partial_5f_l|_{x_5=0}  & = & 0  &  \qquad \Longrightarrow  \qquad &
    d_l^- \, \kappa_{l}^{-}
    & = & -  d_l^+ \, \kappa_{l}^{+}
\\[2ex]
\partial_5f_l|_{x_5=\pi} & = & 0 & \qquad \Longrightarrow  \qquad &
0 & = & \dis \kappa_{l}^{-} \, \sin\left( (\kappa_{l}^{-}-1)\Theta_k(\pi)\right) 
\earr
\]
Clearly, the trivial solution $\kappa_{l}^{-} = 0$, corresponds 
to $m_p = 0$ and $f_l(x_5) = 1$. Other solutions are given by
\beq
\label{flmass}
\sqrt{1+ 4 \frac{m_p^2R_y^2}{c^2}} = \frac{p \pi}{\Theta_k(\pi)} 
\eeq 
with $p \in \mathbb{Z}^+$, thereby quantizing $m_p$.
Similarly, for $f_r$ we have
\[
\barr{rclcrcl}
f_r|_{x_5=0} & = & 0  & \qquad \Longrightarrow  \qquad 
& d_r^+ & = & 
- d_r^-
\\[2ex]
f_r|_{x_5=\pi} & = & 0 & \qquad \Longrightarrow  \qquad & 
       0 & = & \dis \sin\left((\kappa_{r}^{+}+1)\Theta_k(\pi)\right) 
\earr
\]
leading to (as expected) a mass quantization condition identical to that 
for the left-chiral fields. Of course, the $m_p = 0$ state does not exist 
for the $f_r$, implying, in turn, the phenomenologically required 
condition of there being no massless right-handed $SU(2)_L$-doublet field. 

Given a $m_p$, we can now solve the first 
of eqns.(\ref{fermioneqmotion}), to $\ft_{l/r}^{n,p}$ in terms 
of Bessel functions, namely
\begin{equation}
\label{fly}
\barr{rcl}
\ft^{n,p}_l(x_4)& = & \dis e^{c|x_4|/2}\, 
           \left[c_1 J_{\nu_p}(x_{np}e^{c(|x_4|-\pi)}) 
               + c_2 Y_{\nu_p}(x_{np}e^{c(|x_4|-\pi)})  \right]
\\[1ex]
\ft^{n,p}_r(x_4)& = & \dis e^{c|x_4|/2}\,
           \left[ c_3 J_{\nu_p}(x_{np}e^{c(|x_4|-\pi)}) 
                + c_4 Y_{\nu_p}(x_{np}e^{c(|x_4|-\pi)})  \right]
\\[1ex]
\nu_p & \equiv & \dis \sqrt{\frac{1}{4} + \frac{m_p^2 R_y^2}{c^2}}
                      = \frac{p \, \pi}{2 \, \Theta_k(\pi)}
\\
x_{np} & \equiv & \dis M_{np} \frac{R_y}{c} e^{c \pi} \ ,
\earr
\end{equation}
where $c_i$s are the constants of integration. It is interesting to note 
that for $m_p = 0$, we get back the wave functions corresponding to 
the five-dimensional RS theory~\cite{Gherghetta:2000qt}. In particular, 
the massless mode (i.e., $M_{0,0} = 0$ and obtainable only for $m_0 = 0 $) 
has, as expected, a flat profile.

\subsection{KK masses for the fermions}
As we shall see in the next section, the issue of spontaneous symmetry
breaking is a tricky one, for
the inclusion of a bulk Higgs has been
shown~\cite{chang} to resuscitate the hierarchy problem. Localizing
the higgs onto a brane obviates the problem, though.  Given this, we
continue under the assumption of the six-dimensional fields to be
strictly massless and, thus, the only contribution to the masses of
the four-dimensional components would be those due to the
compactification.  Incorporating the effect of the Higgs
field is a straightforward exercise,
and would be undertaken in Sec.\ref{sec:higgs}. 
In any case, with the compactification
scale being much larger than the electroweak scale, the Higgs
contribution would be of little importance to any but the lowest mode
and can be treated as a perturbation.

Determining the spectrum, as usual, needs the imposition of the
boundary conditions. While we already have done so for the $x_5$-modes, the $x_4$-component is
still unrestricted. Noting that the orbifolding demands that the
fermion wavefunctions be even in the $x_4$-direction, these must
satisfy
\beq
\partial_4 \ft^{n,p}_{l}(x_4) |_{x_4=0} = 0 \ , \quad
\partial_4 \ft^{n,p}_{r}(x_4) |_{x_4=\pi} = 0 \ .
\eeq
Rather than attempt to solve for the above for the most general 
choice of $c$ and $k$, we restrict ourselves to the two cases 
that are of relevance in resolving the hierarchy problem, viz. 
small $k$ (large $c$) on the one hand and large $k$ (small $c$) 
on the other.

\subsubsection{Small $k$ and large $c$}
\begin{figure}[!h]
\vspace*{-50pt}
\centerline{
\epsfxsize=8cm\epsfbox{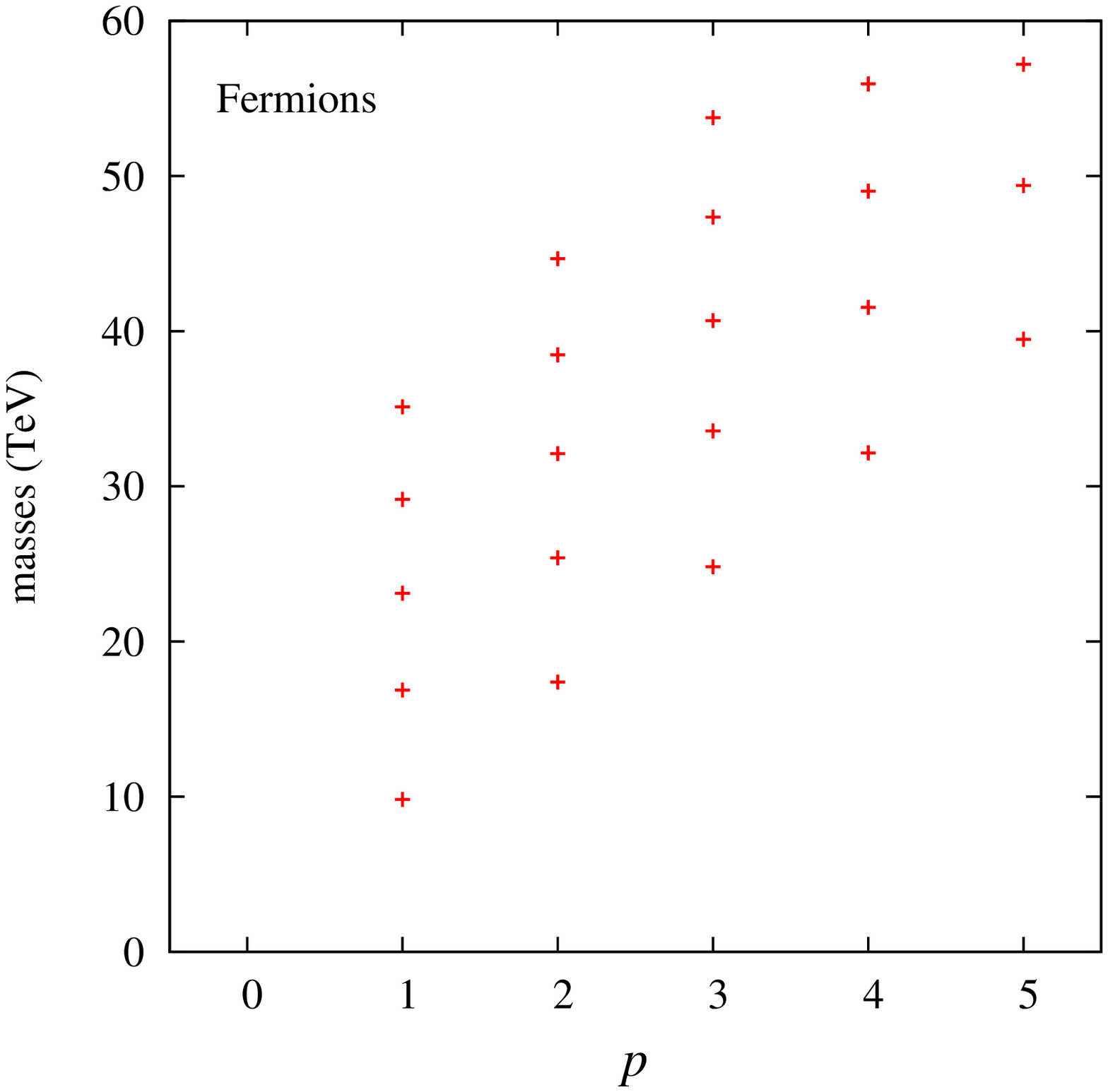}
\epsfxsize=8cm\epsfbox{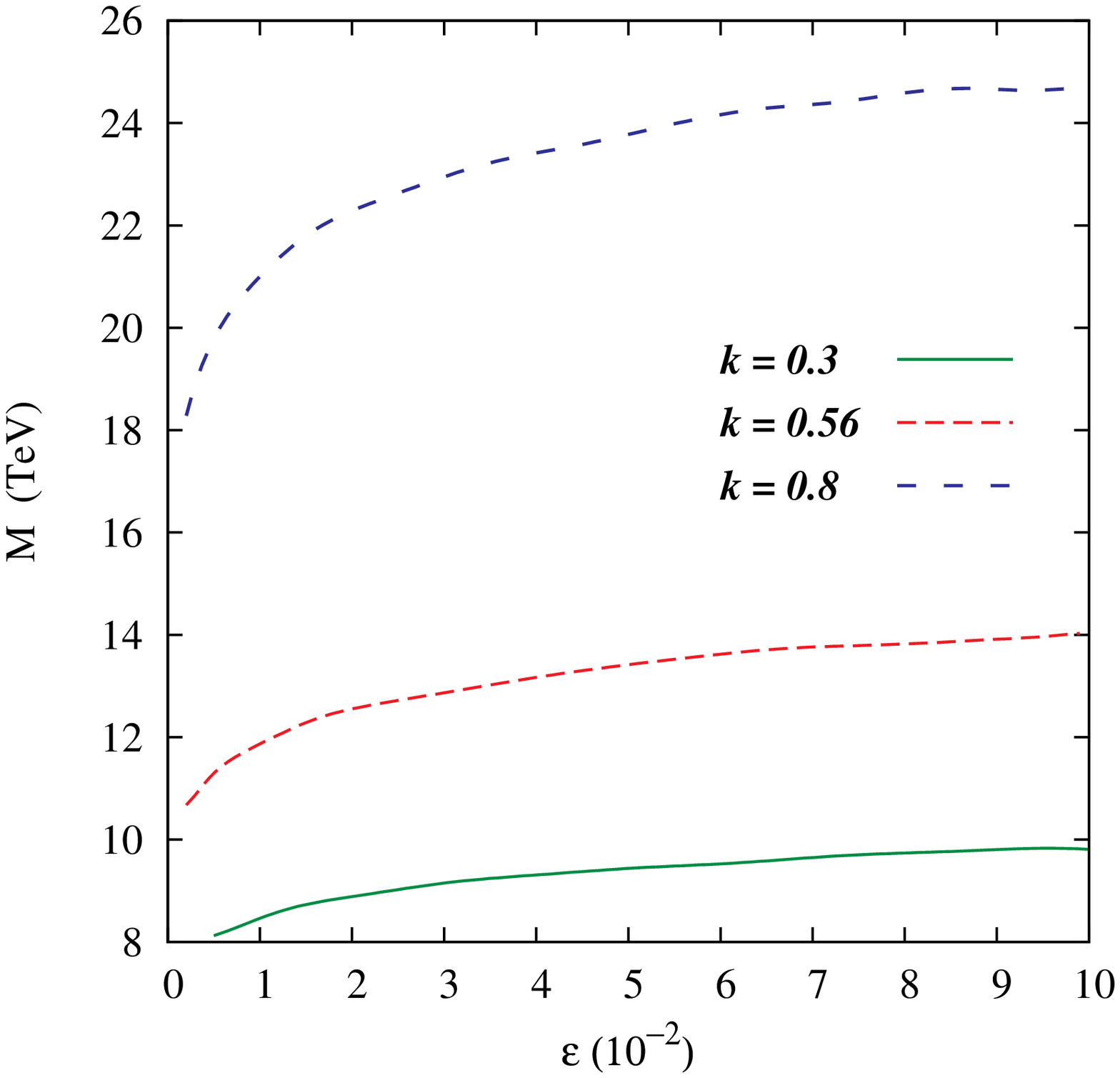}
}
\vspace*{-70pt}
\caption{\em (Left panel) A sample Kaluza-Klein spectrum that satisfies eqn(~\ref{quant_left})
originating from a six-dimensional chiral fermion, for $k=0.3$,
$\alpha = 49$ and $\epsilon = 0.0775$ (see eqn.\ref{RS6_eqns}).  Only the 
first five $n$ levels corresponding to each $p$ are shown.  
(Right panel) The dependence of the mass of the lowest KK mode on $\epsilon$. 
In both the panels,
 $R_y$ set to satisfy the hierarchy eqn(~\ref{hierarchy}).
}
\label{fig:small_kfermion_mass}
\end{figure}

The aforementioned boundary conditions, respectively, give
\begin{equation}
\frac{-c_2}{c_1}  =  
   \frac{e^{-c\pi}x_{np} \, 
       \left[ J_{\nu_{p}-1}(x_{np}e^{-c\pi})
             -J_{\nu_{p}+1}(x_{np}e^{-c\pi})\right] 
          + J_{\nu_{p}}(x_{np}e^{-c\pi})}
        {e^{-c\pi}x_{np} \,
       \left[ Y_{\nu_{p}-1}(x_{np}e^{-c\pi})
             -Y_{\nu_{p}+1}(x_{np}e^{-c\pi})\right]
          +Y_{\nu_{p}}(x_{np}e^{-c\pi})}
\label{xnp_left_1}
\end{equation}
and 
\begin{equation}
\frac{-c_2}{c_1} = \frac{x_{np} \, 
          \left[J_{\nu_{p}-1}(x_{np})-J_{\nu_{p}+1}(x_{np})
          \right] +J_{\nu_{p}}(x_{np})}
       {x_{np} \, \left[ Y_{\nu_{p}-1}(x_{np})
                        -Y_{\nu_{p}+1}(x_{np})\right]
                 +Y_{\nu_{p}}(x_{np})} \ .
\label{xnp_left_2}
\end{equation}
As
$e^{-c\pi}$ is negligibly small in this regime, the corresponding
$Y_{\nu}$ are
very large.  Consequently, in this
regime, the two equations above are simultaneously satisfied only if
\beq
x_{np}\left[J_{\nu_{p}-1}(x_{np})-J_{\nu_{p}+1}(x_{np})\right]
+J_{\nu_{p}}(x_{np}) = 0 \ ,
\label{quant_left}
\eeq
thereby determining the quantized values for $x_{np}$ (for a given
$\nu_p$ determined, in turn, by
eqns.(\ref{fly}\&\ref{flmass})). For a particular choice of the 
parameters, these are exhibited in 
Fig.~\ref{fig:small_kfermion_mass}.

For the right-chiral fields, relations analogous to eqns.(\ref{xnp_left_1}\&\ref{xnp_left_2}) 
would obtain. It is easy to see 
that the corresponding set of $x_{np}$s identically match those in 
eqn.(\ref{quant_left}), but for the zero mode.



\subsubsection{Large $k$ and small $c$}
Since $c$ is now almost infinitesimally small, $a(x_4) \approx 1$ is a very 
good approximation and eqn.(\ref{a11}) can be simplified down to
\[
\partial_4^{2}\ft^{n,p}_{l/r}(x_4) \, = \, -(M_{np}^{2}-m_{p}^{2})R_y ^{2} \ft^{n,p}_{l/r}(x_4) \ .
\] 
This, of course, yields plane wave solutions, which is as expected since 
such a small warping means that the space is essentially flat.
Hence, the spectrum is given by
\[
M_{np}^{2} \approx \frac{n^2}{R_y^2} + m_{p}^{2} \ .
\]
As $R_y$ is very small, we could as well neglect the $n \neq 0$
modes in any discussion of TeV scale physics.
In other words, for all practical
purposes, the fermions act as if they are confined onto a brane.

\section{The Gauge Bosons}
    \label{sec:gauge}
To begin with, we examine the mass spectrum of the gauge bosons postponing
discussion of all interactions until later. Hence, it is convenient to
consider only a $U(1)$ theory in the six dimensional
bulk. Similarly, rather than concerning ourselves with the issue of
symmetry breaking in the bulk, we would introduce an explicit
mass term so as to understand the consequences of 
a bulk mass term.

The gauge boson Lagrangian is given by
\begin{equation}
{\cal L}=\frac{-1}{4}\sqrt{-g}F_{M N}F^{M N} + {\cal L}_{gf} \ .
\end{equation}
While the structure of the gauge-fixing term is, {\em a priori},
undetermined, it is often useful to restrict ourselves to a covariant
choice, so as to manifestly respect the symmetries of the theory,
which, at first sight, 
would seem to be
diffeomorphism invariance. Note, however, that the presence of the
boundary branes (and the orbifolding), actually reduces the symmetry
away from the bulk
to only a four-dimensional general coordinate invariance. 
Hence, we
introduce a curved-space analog of the oft-used (at least in the case of flat extra
dimensions~\cite{Burdman:2005sr}) generalized $R_\zeta$
gauge which, while respecting the four-dimensional general coordinate
invariance, also serves to eliminate the cumbersome kinetic mixing
terms between $A_{\mu}$ and $A_{4,5}$. To be specific, we have
\beq
\barr{rcl}
{\cal L}_{gf} & = & \dis \frac{-\sqrt{-g}}{2 \zeta} \,
     \left[ g^{\mu\nu} \left\{\partial_\mu A_\nu 
                 - \frac{\zeta}{2} \, \left(\Gamma^{4}_{\mu \nu} A_4 
                                      +\Gamma^{5}_{\mu \nu} A_5\right) \right\}
            + \zeta \, (g^{44} D_4 A_4 + g^{55} D_5 A_5) \right]^2
\\[2ex]
& = & \dis
\frac{-R_y r_z b}{2 \zeta} \left[ \eta^{\mu\nu} \partial_{\mu}A_{\nu} 
       + \frac{\zeta}{b} \left( \partial_4 \frac{a^2 b A_4}{R_y^2} 
                              + \partial_5 \frac{a^2 b^3 A_5}{r_z^2} \right)
           \right]^2 \ .
\earr
\eeq
This does not exhaust all of the gauge symmetries.  On
compactification down to four dimensions, the components $A_{4,5}$
would, naturally, give rise to a `tower of towers' of scalars
transforming under the adjoint representation of the gauge group. On
the other hand, the very act of the higher modes of $A_\mu$ becoming
massive could be viewed as the result of a Higgs mechanism wherein the
adjoint Goldstone has been absorbed. In other words, in the 
unitary gauge,
only one linear combination of $A_{4,5}$ may survive. But as in the
case with UED, this conclusion
is dependent on the gauge choice. In particular,
for $\zeta = 1$, both sets of the scalars survive.
Adopting, for the time, the Minkowski metric, the quadratic term for the 
vector field is now given by 
\beq
{\cal L}_{A_{\mu}} = \frac{-R_y r_z }{2} \,
    \left[ b A_{\kappa} (-\partial^2 \eta^{\kappa \lambda}
                         +\partial^{\lambda}\partial^{\kappa}) A_{\lambda} 
         + \frac{a^2b}{R_y^2} (\partial_4 A_\kappa) (\partial_4 A^{\kappa})
         + \frac{a^2b^3 }{r_z^2} (\partial_5 A_\kappa) (\partial_5 A^\kappa)
    \right] \ ,
\eeq
while for the adjoint scalars, with field redefinitions 
\beq
\tilde{A}_{4} \equiv \sqrt{\frac{r_z}{R_y}} \; A_{4} \ , 
\qquad {\rm and} \qquad
\tilde{A}_{5} \equiv \sqrt{\frac{R_y}{r_z}} \; A_{5}\ ,
\eeq
 it is
\beq
{\cal L}_{\tilde A_4} = \frac{-1}{2}\left[
      a^2b(\partial_\mu \tilde{A}_4) \, (\partial^\mu \tilde{A}_4) 
      + \frac{a^4 b^3}{r_z^2}(\partial_5 \tilde{A}_4\big)^2
      + \frac{1}{R_y^2 b} \left\{ \partial_4 (a^2b\tilde{A}_4)\right\}^2\right] \ ,
\eeq
\beq
{\cal L}_{\tilde A_5} = \frac{-1}{2}\left[
      a^2b^3(\partial_\mu \tilde{A}_5) \, (\partial^\mu \tilde{A}_5) 
      + \frac{a^4 b^3}{R_y^2}(\partial_4 \tilde{A}_5\big)^2
      + \frac{1}{r_z^2 b} \left\{ \partial_5 (a^2b^3\tilde{A}_5)\right\}^2\right] \ ,
\eeq
and reminiscent of the action for a scalar field. There also exists a 
mixing term
\beq
{\cal L}_{\rm mix} = 2\frac{1}{R_y r_z} a^3 a' \tilde A_4 \partial_5(b^3 \tilde A_5) \ ,
\eeq
and, hence, ideally, one
should rediagonalize the $\tilde A_4$---$\tilde A_5$ system. However,
as a zeroth approximation, one may neglect ${\cal L}_{\rm mix}$
altogether and derive the wavefunctions and propagators from ${\cal
L}_{\tilde A_4}$ and ${\cal L}_{\tilde A_5}$ alone. Using the thus
derived wavefunctions in ${\cal L}_{\rm mix}$, it can be seen that
this term is numerically rather subdominant. Allied with the fact that
the zero modes of $\tilde A_{4,5}$ would not survive and only the
heavy KK-modes come into play, the neglect of ${\cal L}_{\rm mix}$ has
almost no discernible consequence as far as low energy phenomenology
is concerned. We shall, thus, continue with this approximation, while
bearing in mind that ${\cal L}_{\rm mix}$ ought to be taken into
account when dealing with subleading effects as also in the context of
questions such as unitarity.

To the quadratic kinetic terms, we may 
add a mass term (presumably originating from spontaneous symmetry 
breaking in the bulk), viz.
\beq
{\cal L_M} = -\, \frac{\sqrt{-g}}{2}M^2A_{N}A^{N} \ . 
\eeq
We may now express the field in terms of the eigenstates of the
extra-dimensional parts of the aforementioned
differential operators.
These are but the analogues of the
``plane wave'' solutions and given by
\beq
\barr{rcl} 
A_\kappa & = & \dis 
\frac{1}{\sqrt{R_y r_z}} \sum_{n,p} 
A^{(n,p)}_\kappa(x^{\mu}) \, \eta_{n,p}(x_4) \, \chi_p(x_5) 
\\[1ex]
\tilde{A}_{4} & = & \dis 
\sum_{n,p} \tilde{A}^{(n,p)}_{(4)}(x_{\mu}) \, \eta^{(4)}_{n,p}(x_4) \, \chi^{(4)}_{p}(x_5) \ ,
\\[1ex]
\tilde{A}_{5} & = & \dis 
\sum_{n,p} \tilde{A}^{(n,p)}_{(5)}(x_{\mu}) \, \eta^{(5)}_{n,p}(x_4) \, \chi^{(5)}_{p}(x_5) \ ,
\earr
    \label{gauge_boson_comp}
\eeq
with the components satisfying the orthogonality relations
\beq
\label{vector:ortho}
\barr{rcl}
\dis \int dx_5 \, b(x_5) \, \chi_p(x_5)\chi_{p'}(x_5) & = & \dis \delta_{p p'}
\\[1ex]
\dis \int dx_4 \, \eta_{n,p}(x_4) \, \eta_{n',p'}(x_4) 
     & = & \dis \delta_{n n'} \, \delta_{p p'}
\\[1ex]
\dis \int dx_5 \,b(x_5) \, \chi^{(4)}_p(x_5)\chi^{(4)}_{p'}(x_5) & =& \dis 
\delta_{p p'}
\\[1ex]
\dis \int dx_4 \, a^2(x_4) \,\eta^{(4)}_{n,p}(x_4) \, \eta^{(4)}_{n',p'}(x_4)
     & = & \dis \delta_{n n'} \, \delta_{p p'} 
\\[1ex]
\dis \int dx_5 \,b^3(x_5) \, \chi^{(5)}_p(x_5)\chi^{(5)}_{p'}(x_5) & =& \dis 
\delta_{p p'}
\\[1ex]
\dis \int dx_4 \, a^2(x_4) \,\eta^{(5)}_{n,p}(x_4) \, \eta^{(5)}_{n',p'}(x_4)
     & = & \dis \delta_{n n'} \, \delta_{p p'} \ .
\earr
\eeq
The corresponding equations of motion for the vector modes are 
\begin{equation}
\label{eqofmotionz}
\barr{rcl}
\dis \frac{1}{r_z^2}\partial_5(b^3 \, \partial_5 \chi_p) \, -M^2 \, b^3 \, 
      \chi_p  & = & - m_p^2 \, b \, \chi_p
\\
\dis \frac{1}{R_y^2}\partial_4(a^2\partial_4\eta_{n,p}) - m_p^2 \, a^2 \, \eta_{n,p} & =  & \dis
   - m_{np}^2 \eta_{n,p} \ ,
\earr
\end{equation}
whose solutions could be written as 
\beq
\barr{rcl}
\chi_p(x_5) & = & \dis \frac{1}{\bar{B}_p} {\rm sech}^{3/2}(k x_5) \,
      \left( c_1 \, P_{\nu_p}^{u/2}(\tanh k x_5) + 
             c_2 \, Q_{\nu_p}^{u/2}(\tanh k x_5)\right)
\\[2ex]
\eta_{n,p}(x_4) & = & \dis 
    \frac{e^{c|x_4|}}{B_{np}} \Big( J_{\nu_n}(y_n) + c_{np} Y_{\nu_n}(y_n)\Big)
\\[2ex]
y_n & \equiv & \dis m_{np}\frac{r_z}{k} e^{c|x_4|} \, \cosh(k \pi) = 
                    m_{np} \, \frac{R_y}{c} \, e^{c|x_4|} 
\\[2ex]
\nu_n & = & \dis \sqrt{1+\frac{r_z^2}{k} \, m_p^2 \, \cosh^2(k \pi)}  
\\[2ex]
\nu_p & = & \dis \frac{-1}{2} + \nu_n
\\[2ex]
u & = & \dis \sqrt{9+\frac{4M^2r_z^2}{k^2}} \ ,
\earr
    \label{gauge_soln}
\eeq
where $c_{1,2}$ (we explicitly retain both as this is useful
in studying the boundary conditions) and $c_{np}$
are arbitrary constants while $\bar B_p$ and $B_{np}$ provide the
normalization. The associate
Legendre functions, appearing also in the
description of the fermions, or at any stage of the 
six-dimensional
theory, are reminiscent of the $x_5$-dependence of the graviton
wavefunctions~\cite{Arun:2014dga} and are a feature of the nested warping.

The equations of motion for the adjoint scalar $A_4$ are
\begin{equation}
 \label{eqofmotionyA4}
\barr{rcl}
\dis \frac{1}{r_z^2}\partial_5\left(b^3\partial_5( \chi^{(4)}_{p})\right)  
-  M^2 b^3 \chi^{(4)}_{p} & = & \dis - \tilde m_p^2 \, b \, \chi^{(4)}_{p}
\\[2ex]
\dis \frac{1}{R_y^2}\partial_4(\partial_4a^2 \eta^{(4)}_{n,p}) - a^2 \tilde m_{p}^2 \eta^{(4)}_{n,p} 
   & = & \dis - \tilde m_{np}^2 \eta^{(4)}_{n,p} \ ,
 \earr
\end{equation}
leading to
\beq
\barr{rcl}
\chi^{(4)}_{p}(x_5) & = & \dis 
   \frac{1}{\bar{{\cal E}}_p} {\rm sech}^{3/2}(k x_5) \, 
      \left[ s_1 \, P_{\tilde\nu_p}^{\tilde v/2}(\tanh k x_5) 
           + s_2 \, Q_{\tilde\nu_p}^{\tilde v/2}(\tanh k x_5) \right] 
\\[2ex]
\eta^{(4)}_{n,p}(x_4) & = & \dis \frac{1}{{\cal E}_{np}}e^{2c|x_4|} \,
   \left[ J_{\tilde\nu_n}(y_n) + s_{np} Y_{\tilde\nu_n}(y_n)\right]
\\[2ex]
\tilde\nu_n & = & \dis \sqrt{\frac{r_z^2}{k^2} \, \tilde m_{p}^2 \cosh^{2}(k \pi)}
\\[2ex]
\tilde\nu_p & = & \dis \frac{-1}{2}+\sqrt{1+\frac{r_z^2}{k^2} \tilde m_p^2 \cosh^{2}(k \pi)}  =  \nu_p
\\[2ex]
\tilde{v} & = & \dis \sqrt{9+\frac{4M^2r_z^2}{k^2}}  =  u \ .
\earr
    \label{adjoint_soln4}
\eeq
where $s_{1,2}$ and $s_{np}$ are constants of integration, while 
${\cal E}_{np}$ and ${\cal E}_{p}$ serve to normalize.
Similarly, the equations of motion for the adjoint scalar $A_5$ are seen to be
\begin{equation}
\label{eqofmotionyA5}
\barr{rcl}
\dis \frac{1}{r_z^2}\partial_5\left(b^{-1}\partial_5( b^3 \chi^{(5)}_{p})\right)  
-  M^2 b^2 \chi^{(5)}_{p} & = & \dis - \bar m_p^2 \, \chi^{(5)}_{p}
\\[2ex]
\dis \frac{1}{R_y^2}\partial_4(a^4\partial_4\eta^{(5)}_{n,p}) - a^4 \bar m_{p}^2 \eta^{(5)}_{n,p} 
   & = & \dis - \bar m_{np}^2 a^2 \eta^{(5)}_{n,p} \ ,
\earr
\end{equation}
leading to
\beq
\barr{rcl}
\chi^{(5)}_p(x_5) & = & \dis 
   \frac{1}{\bar{{\cal D}}_p} {\rm sech}^{5/2}(k x_5) \, 
      \left[ d_1 \, P_{\bar\nu_p}^{\bar v/2}(\tanh k x_5) 
           + d_2 \, Q_{\bar\nu_p}^{\bar v/2}(\tanh k x_5) \right] 
\\[2ex]
\eta^{(5)}_{n,p}(x_4) & = & \dis \frac{1}{{\cal D}_{np}}e^{2c|x_4|} \,
   \left[ J_{\bar\nu_n}(y_n) + d_{np} Y_{\bar\nu_n}(y_n)\right]
\\[2ex]
\bar\nu_n & = & \dis \sqrt{4+\frac{r_z^2}{k^2} \, \bar m_{p}^2 \cosh^{2}(k \pi)}
\\[2ex]
\bar\nu_p & = & \dis \frac{-1}{2}+\sqrt{1+\frac{r_z^2}{k^2} \bar m_p^2 \cosh^{2}(k \pi)} = \nu_p
\\[2ex]
\bar{v} & = & \dis \sqrt{1+\frac{4M^2r_z^2}{k^2}} \ .
\earr
    \label{adjoint_soln5}
\eeq
Once again, $d_{1,2}$ and $d_{np}$ are constants of integration, while 
$\bar{\cal D}_p$ and ${\cal D}_{np}$ provide normalizations.
With these conditions in place, the quadratic part 
of the Lagrangian for the vector fields can be expressed 
in terms of the KK-towers as 
\[
\barr{rcl}
{\cal L}_{A_{\mu}} & = & \dis \sum_{n,p} \left[
   \frac{-1}{4}F^{(n,p)}_{\mu \nu}F^{\mu \nu (n,p)} 
     - \, \frac{1}{2}  m_{np}^{2} A^{(n,p)}_{\mu}A^{\mu (n,p)} - \frac{1}{2} (\partial_\mu A^{\mu(n,p)})^2 \right] \ , 
\earr
\]
while, for the adjoint scalars, we have
\[
{\cal L}_{A_5} \, = \, -\frac{1}{2}\big(\partial_\mu \tilde{A}_{4}^{(n,p)}\big)^2 - \frac{1}{2} \tilde m_{np}^2\tilde{A}_{4}^{(n,p)2} \ .
\]
and
\[
{\cal L}_{A_5} \, = \, -\frac{1}{2}\big(\partial_\mu \tilde{A}_{5}^{(n,p)}\big)^2 - \frac{1}{2} \bar m_{np}^2\tilde{A}_{5}^{(n,p)2} \ .
\]

\subsection{Gauge and Adjoint Scalar Masses}
\label{mass spectra}
Our aim, now, is to compute the allowed values for $m_{np}$ , $\tilde
m_{np}$ and $\bar m_{np}$.  In each case, one must first find the
$x_5$-equation eigenvalues ($m_p$, $\tilde m_p$ and $\bar m_p$
respectively) using the boundary conditions for the corresponding
wavefunctions and, then, find the desired spectrum in terms of
these. For ease of appreciation, we perform the exercise in the
reverse order. We first establish the general conditions and then
examine the situation for the two particular cases of interest (in
terms of the relative sizes of $k$ and $c$).  

In doing so, it
should be borne in mind that the bulk mass term $M$, that we have
considered until now, would identically disappear, to be replaced, in
the electroweak sector, by the spontaneous breaking term. The latter,
in all our constructions, would be confined to a brane, and its {\em
effective scale} would, naturally, turn out to be the electroweak
scale to be compared with the much larger compactification scales
$R_y^{-1}$ or $r_z^{-1}$. Thus, it stands to reason that the effect of 
the spontaneous breaking term in the spectrum and the wavefunctions would be 
negligible, except, perhaps, for the ground state.
Indeed, as the experience with the RS case~\cite{csabaeffective} 
has shown,  its role there too is subdominant. Consequently, we will 
postpone a discussion of such terms until Sec.\ref{sec:bulk_mass}.

\subsubsection{Boundary conditions in the $x_4$--direction}

As of now, we have not considered the existence of any localized
fields\footnote{The situation would change when we introduce the Higgs
  field and we shall explicitly take this into consideration in
  Sec.\ref{sec:higgs}.}.  Consequently, the wavefunctions
$\eta_{n,p}(x_4)$ must be differentiable, especially at the ends 
of the world. This implies that 
\begin{equation}
\label{ycond}
\dis 
- c_{np}
= \frac{ \alpha_{np} \, e^{c \, (|x_4| \, - \, \pi)} 
             J_{\nu_p -\frac{1}{2}}(\alpha_{np} e^{c \, (|x_4| \, - \, \pi)})
        - (-\frac{1}{2} + \nu_p) J_{\nu_p +\frac{1}{2}}(\alpha_{np} e^{c \, (|x_4| \, - \, \pi)})}
       { \alpha_{np} e^{c \, (|x_4| \, - \, \pi)} 
        Y_{\nu_p - \frac{1}{2}}(\alpha_{np} e^{c \, (|x_4| \, - \, \pi)})
-(-\frac{1}{2} + \nu_p) Y_{\nu_p +\frac{1}{2}}(\alpha_{np} e^{c \, (|x_4| \, - \, \pi)})}
\end{equation}
for each of $x_4 = 0, \pi$. Here, we have defined 
\begin{equation}
\alpha_{np} \equiv m_{np} \, \frac{R_y}{c} \, e^{c \, \pi} \ .
\end{equation}
Once $\nu_p$ is known, the two conditions of 
eqn.(\ref{ycond}), together, 
determine $\eta_{n,p}(x_4)$ as well as serve to quantize 
$\alpha_{np}$ (and, hence, $m_{np}$).

Similarly, since $a(x_4)^2\eta^{(4)}_{n,p}(x_4)$ and $\eta^{(5)}_{n,p}(x_4)$ are
even functions, their derivatives have to vanish at both the fixed
points ($x_4 = 0, \pi$) and this gives
\begin{equation}
 s_{np}
= \, - \,\frac{J_{\tilde\nu_n -1}(\tilde\alpha_{np} e^{c \, (|x_4| \, - \, \pi)})
        - J_{\tilde\nu_n +1}(\tilde\alpha_{np} 
                     e^{c \, (|x_4| \, - \, \pi)})}
       { Y_{\tilde\nu_n -1}(\tilde\alpha_{np} e^{c \, (|x_4| \, - \, \pi)})
-Y_{\tilde\nu_n+1 }(\tilde\alpha_{np} e^{c \, (|x_4| \, - \, \pi)})}\Big|_{x_4 = 0, \pi}
\label{ycondgamma4}
\end{equation}
and
\begin{equation}
 d_{np}
= \, - \, \frac{ \bar\alpha_{np} \, e^{c \, (|x_4| \, - \, \pi)} 
             J_{\bar\nu_n +1}(\bar\alpha_{np} e^{c \, (|x_4| \, - \, \pi)})
        - (2 + \bar\nu_n) J_{\bar\nu_n }(\bar\alpha_{np} 
                     e^{c \, (|x_4| \, - \, \pi)})}
       { \bar\alpha_{np} e^{c \, (|x_4| \, - \, \pi)} 
        Y_{\bar\nu_n +1}(\bar\alpha_{np} e^{c \, (|x_4| \, - \, \pi)})
-(2 + \bar\nu_n) Y_{\bar\nu_n }(\bar\alpha_{np} e^{c \, (|x_4| \, - \, \pi)})}\Big|_{x_4 = 0, \pi}
\label{ycondgamma5}
\end{equation}
for $\eta^{(4)}_{n,p}$ and $\eta^{(5)}_{n,p}$ respectively. Here,
\beq
\tilde \alpha_{np} = \tilde m_{np} \, \frac{R_y}{c} \, e^{c \, \pi} 
\qquad {\rm and} \qquad 
\bar \alpha_{np} = \bar m_{np} \, \frac{R_y}{c} \, e^{c \, \pi} \ .
\eeq
Again, once $\tilde{\nu}_n$ and $\bar\nu_n$ is known, the two conditions in
eqn.(\ref{ycondgamma4}) and eqn.(\ref{ycondgamma5}) together determine $\eta^{(4,5)}_{n,p}(x_4)$ as well
as quantize the masses.

\subsubsection{Boundary conditions in the $x_5$--direction}
\label{sec:boundary_gauge_gen}

As $\chi_p(x_5)$ are even functions of $x_5$, their derivatives would vanish at 
$x_5 = 0$. This translates to 
\begin{equation}
\cot \theta_p \equiv \frac{c_1}{c_2} = \frac{-\pi}{2} \, \cot \frac{\pi \, (\nu_p + u/2)}{2} \ .
\end{equation}
An analogous condition would be obtained for the derivative 
at $x_5 = \pi$, but that is best analyzed separately for small and large $k$,
which we come to later.

We have already seen that since $\eta^{(4,5)}_{n,p}(x_4)$
are even functions, they need to satisfy Neumann boundary conditions at 
$x_4 = 0, \pi$. Thus, the absence of massless adjoint scalars necessitates 
that we impose Dirichlet boundary
condition (a consequence of the orbifolding)
on $\chi^{(4,5)}_p(x_5)$, namely 
$\chi^{(4,5)}_p(0) = 0 = \chi^{(4,5)}_p(\pi)$. 
This, then, implies that
\[
\frac{- s_2}{s_1} 
= \frac{P_{\tilde\nu_p}^{\tilde v/2}(0)}{Q_{\tilde\nu_p}^{\tilde v/2}(0)} 
 =  \frac{P_{\tilde\nu_p}^{\tilde v/2}(\tanh k\pi)}{Q_{\tilde\nu_p}^{\tilde v/2}(\tanh k\pi)} \, \,  , \, \,
\frac{- d_2}{d_1}
= \frac{P_{\bar\nu_p}^{\bar v/2}(0)}{Q_{\bar\nu_p}^{\bar v/2}(0)} 
 =  \frac{P_{\bar\nu_p}^{\bar v/2}(\tanh k\pi)}{Q_{\bar\nu_p}^{\bar v/2}(\tanh k\pi)} \ .
\]
In other words, the eigenvalue spectrum is given by
\beq
P_{\tilde\nu_p}^{\tilde v/2}(0) \, Q_{\tilde\nu_p}^{\tilde v/2}(\tanh k\pi)
- Q_{\tilde\nu_p}^{\tilde v/2}(0) \, P_{\tilde \nu_p}^{\tilde v/2}(\tanh k\pi) = 0 \
   \label{z_bc_adj4}
\eeq
for $A_4$ and
\beq
P_{\bar\nu_p}^{\bar v/2}(0) \, Q_{\bar\nu_p}^{\bar v/2}(\tanh k\pi)
- Q_{\bar\nu_p}^{\bar v/2}(0) \, P_{\bar \nu_p}^{\bar v/2}(\tanh k\pi) = 0 \
   \label{z_bc_adj5}
\eeq
for $A_5$.

At this point, it is worthwhile to remember our earlier discussion about the
limit of vanishing bulk mass ($M = 0$). This immediately leads to  
$\bar v = 1$ in eqns.(\ref{adjoint_soln5}). Now, if we look for $\bar m_p = 0$
(for the $\tilde A_5$ spectrum), then we need to concentrate on $\bar \nu_p = 1/2$ 
in eqn.(\ref{z_bc_adj5}) above, or, 
\[
P_{1/2}^{1/2}(0) \, Q_{1/2}^{1/2}(\tanh k\pi)
- Q_{1/2}^{1/2}(0) \, P_{1/2}^{1/2}(\tanh k\pi) = 0 \ .
\]
However, since the function $P_{1/2}^{1/2}(x) / Q_{1/2}^{1/2}(x)$ is
monotonic, this equation can never be satisfied for any $k$. 
In other words, $\bar
m_p = 0$ is strictly disallowed. Equivalently, not only is the 
unwanted zero mode $A_5^{(0,0)}(x_\mu)$ absent, but all the modes $\tilde
A_5^{(n,0)}(x_\mu)$ do not exist\footnote{This is reminiscent of the 
fermion spectrum.}. 

Similarly, for the $\tilde A_4$ spectrum, $M = 0$ would 
imply $\tilde v = 3$, leading to the $Q_{1/2}^{\tilde v / 2}$ vanishing 
identically. The imposition of the aforementioned boundary condition 
then implies that the corresponding $\chi^{(4)}_0(x_5)$ must 
vanish identically too. Thus, once again, the 
requirement that there be no massless $\chi^{(4)}$ scalar has the 
consequence that the entire putative tower comprising 
of the modes $\tilde A_4^{(n,0)}(x_\mu)$ disappears identically.

For both cases, the argument is quite robust and carries 
through even in the presence of brane-localized spontaneous symmetry 
breaking term. 

As for the rest of the spectrum, this, along with that for 
the gauge 
boson excitations, is 
best analyzed separately for large and small $k$, and this we come to next.

\subsubsection{Small $k$ and large $c$}
Since the Legendre functions are well-behaved in this domain, we could
use the Neumann boundary conditions for $\chi_p(x_5)$ at $x_5 = \pi$ in a straightforward
fashion, and this implies
\begin{equation}
\barr{rcl}
0 & = & \dis 
\cot\theta_p \, (1-2\nu_p) \tau_\pi \, P_{\nu_p}^{u/2}(\tau_\pi)
+\cot\theta_p\,(2+2\nu_p-u) \, P_{\nu_p +1}^{u/2}(\tau_\pi)
\\
&+& \dis (1-2\nu_p) \tau_\pi \, Q_{\nu_p \
}^{u/2}(\tau_\pi)+(2+2\nu_p-u) \, Q_{\nu_p +1}^{u/2}(\tau_\pi) 
\earr
\end{equation}
where $\tau_\pi \equiv \tanh( k \, \pi)$. This equation has to be
solved numerically to obtain the discrete set of values allowed to $\nu_p$.

For a given $\nu_p$, to solve for $m_{np}$, we need to consider both
of eqns.(\ref{ycond}).  Note that, for the $(k, c)$ values under
discussion, the combination 
$R_y^{-1} \, e^{-c \pi}$
 roughly gives the
electroweak scale~\cite{Arun:2014dga}. Thus, if $m_{np}$ are to be
important in low-energy phenomenology, $\alpha_{np} \lapp{\cal O}(1)$.
On the other hand, with $e^{c \, \pi}$ being very large, the argument
of the Bessel functions essentially vanishes at $x_4 = 0$. This implies that
$c_{np} \approx 0$, and using this in
the boundary condition at $x_4 = \pi$, we have
\begin{equation}
2 \alpha_{np} J_{\nu_p - 1/2}(\alpha_{np} ) + (1 - 2 \nu_p )
J_{\nu_p +1/2}(\alpha_{np} )  =  0  \ .
    \label{gauge_KK}
\end{equation}
Solving this, for a given $\nu_p$, would lead to
the gauge boson KK tower starting with $A_{\mu}^{(1,0)}$. 

Let us, now, turn our attention to the adjoint scalars.
Quite similar to the case for the vector modes, 
the two eqns.(\ref{ycondgamma4}) are satisfied only if $s_{np} = 0$ and, hence, 
\begin{equation}
\label{A4masseqaution}
J_{\tilde \nu_n-1}(\tilde \alpha_{np} ) - \,
        J_{\tilde \nu_n +1}(\tilde \alpha_{np} ) =  0 \ .
\end{equation}
Similarly, the
two eqns.(\ref{ycondgamma5}) are satisfied only if $d_{np} = 0$ and, hence, 
\begin{equation}
\label{A5massequation}
\bar \alpha_{np}  J_{\bar\nu_n+1}(\bar \alpha_{np} ) - (2 + \bar\nu_n) \,
        J_{\bar\nu_n }(\bar\alpha_{np} ) =  0 \ .
\end{equation}
In either case, no solution exists for $n = 0$, and, thus,
the first nonzero components of the 
adjoint scalar fields would be $A_{4}^{(0,1)}(x_{\mu})$ 
and $A_{5}^{(0,1)}(x_{\mu})$.

\begin{figure}[!h]
\vspace*{-50pt}
\epsfxsize=8cm\epsfbox{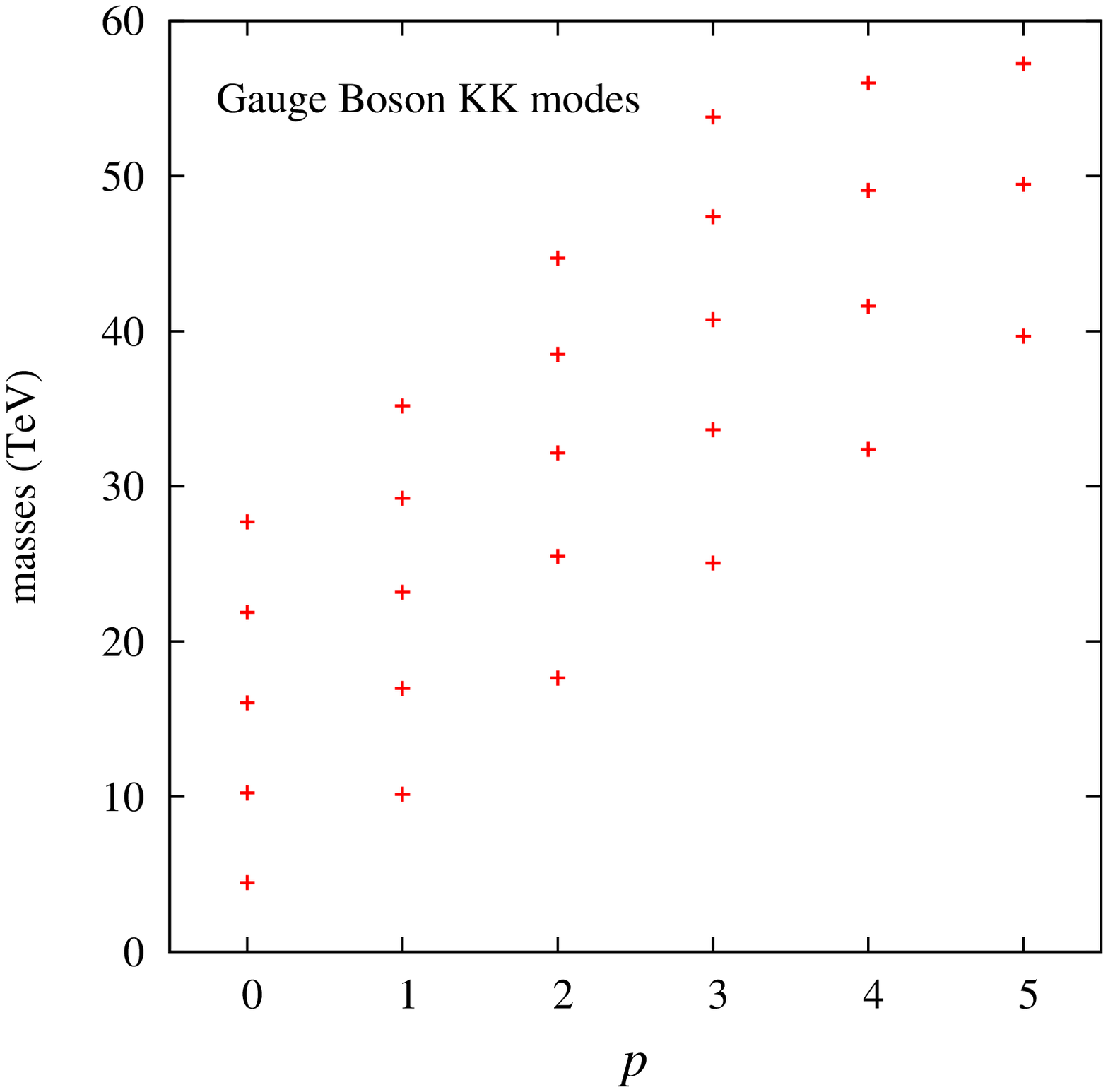}
\epsfxsize=8cm\epsfbox{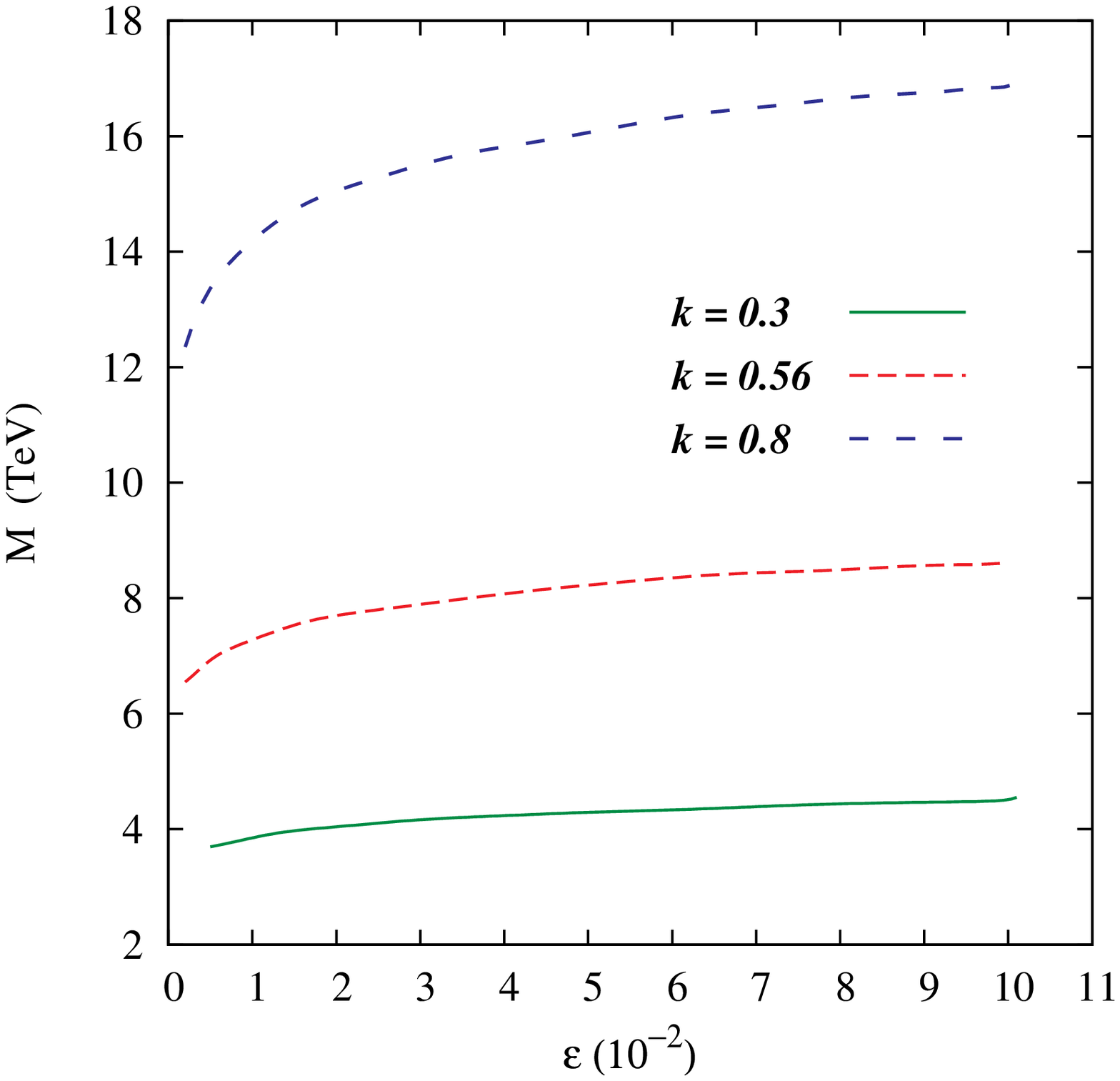}
\vspace*{-70pt}
\caption{\em (Left panel)The spectrum
 of vector gauge bosons with zero
  bulk mass for k=0.3, $\alpha = 49$ and $\epsilon = 0.0775$. Only the
  first five $n$ levels corresponding to each $p$ are shown. (Right
  panel) The dependence of the mass of the lowest KK mode on $\epsilon$.
  In both the panels, $R_y$ set to satisfy the hierarchy
  eqn(~\ref{hierarchy}).  }
  \label{fig:small_k_gaugemass1}
  \end{figure}

\begin{figure}[!h]
\vspace*{-50pt}
\centerline{
\epsfxsize=8cm\epsfbox{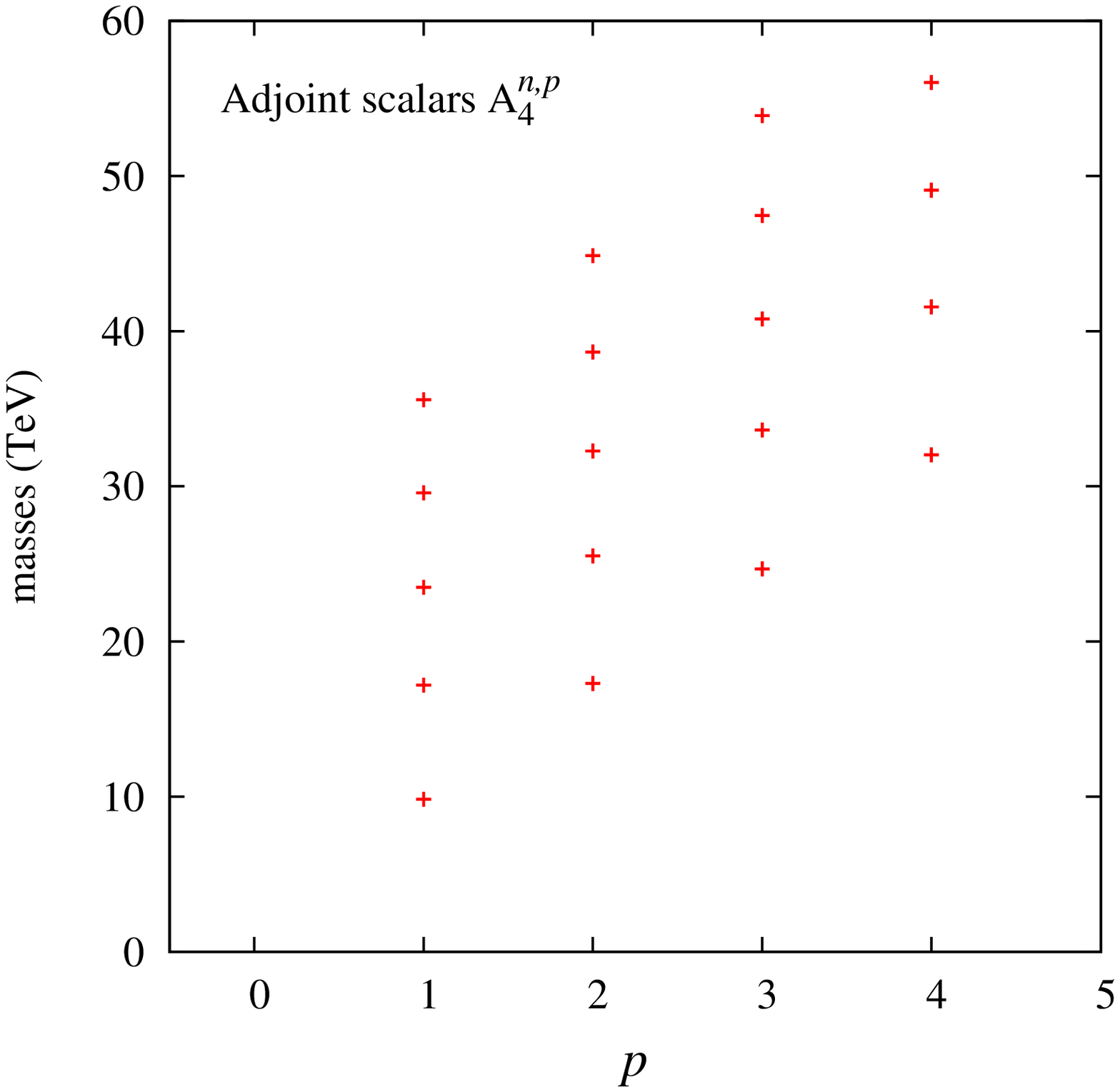}
\epsfxsize=8cm\epsfbox{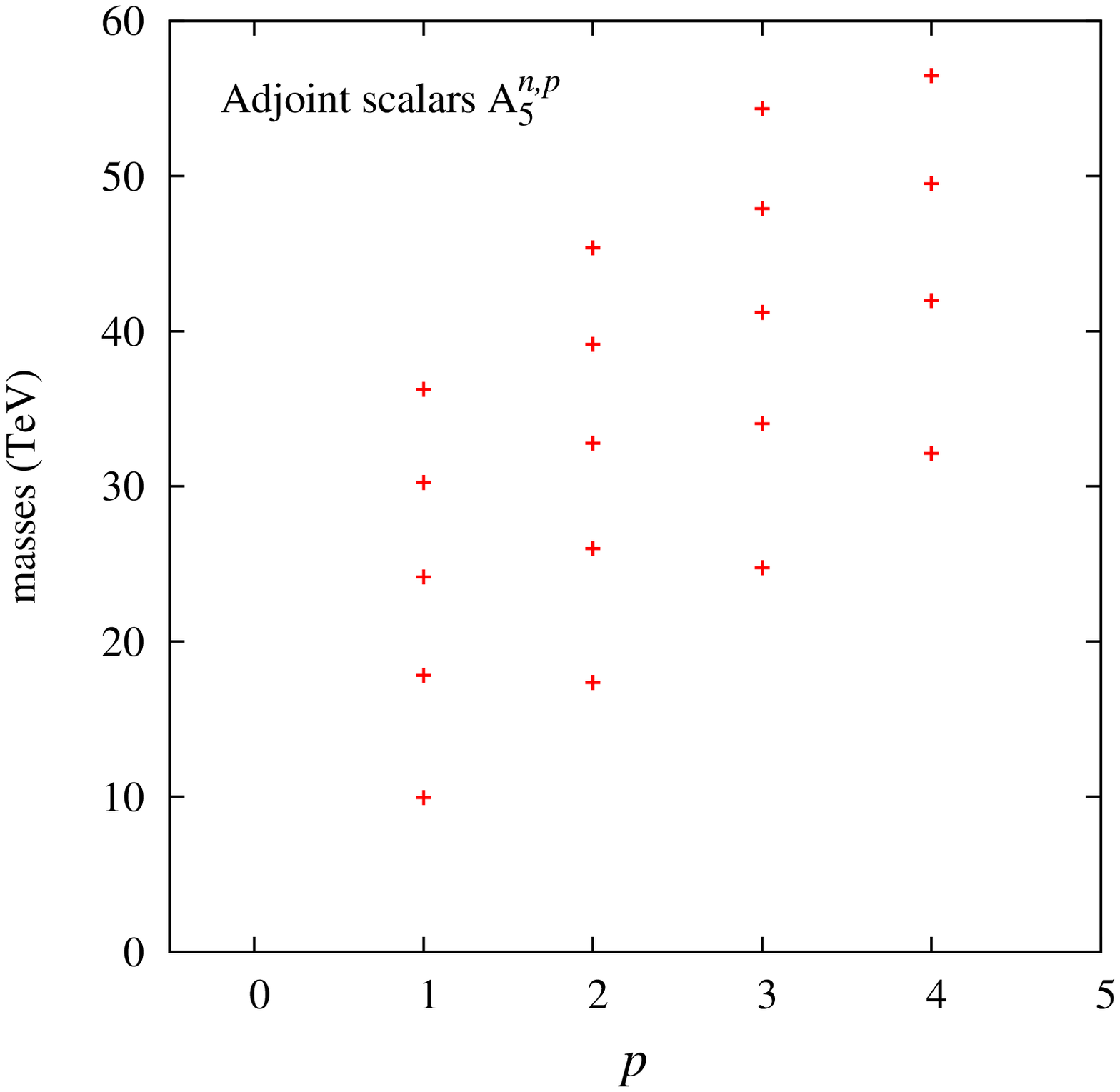}
}
\vspace*{-70pt}
\caption{\em As in the left panel of
  Fig.~\ref{fig:small_k_gaugemass1}, but for the adjoint scalars
  $\tilde A_4^{(n,p)}$ (left panel) and $\tilde A_5^{(n,p)}$ (right
  panel) instead.}
\label{fig:small_k_gaugemass2}
\end{figure}

The quantized masses that satisfy the
equations \ref{gauge_KK}, \ref{A4masseqaution} and
\ref{A5massequation} are shown in Figs. \ref{fig:small_k_gaugemass1}
and \ref{fig:small_k_gaugemass2}.

\subsubsection{Large $k$ and small $c$}
Denoting $\tau = \tanh(k \, x_5)$, we may re-express the wavefunctions 
$\chi_p(x_5)$ as 
\[
\chi_p(\tau) = \frac{1}{N_p}(1-\tau^2)^{\frac{3}{4}} \,
     \left[ \cot\theta_p \, P_{\nu_p}^{u/2}(\tau) + Q_{\nu_p}^{u/2}(\tau)\right] \ .
\]
For  $x_5 = \pi$, we are very close to $\tau = 1$. 
Defining $f(\tau) \equiv d \chi_p / d \tau$, we then have
\[
f(\tau)\Big|_{\tau=1-\epsilon} = \frac{1}{2 \sqrt{2\pi}}
     (2\nu_p-1)(3+2\nu_p) \cot\theta_p + {\cal O}(\epsilon) \ ,
\]
which automatically 
vanishes for the zero mode as the corresponding $\nu_p = 1/2 $. 
For the higher modes to satisfy the Dirichlet boundary condition, 
we require $\cot \theta_p \, = \, 0$, or in
other words $\nu_p  =  2\, p -1/2$, where 
$p \in \mathbb{Z}^+$.

Since we are now in the very small $c$ regime, the warping in the $x_4$ 
direction is virtually nonexistent, and the modes with $ n \neq 0 $ will 
have a mass $\sim n \, R_y^{-1}$ and, hence, will decouple from the theory. 
This is exactly analogous to the case for the corresponding graviton
modes~\cite{Arun:2014dga}. 

The boundary conditions for the adjoint scalars simplifies to
\[
 Q_{\tilde\nu_p}^{\tilde v/2}(0) \, P_{\tilde\nu_p}^{\tilde v/2} (\tanh k\pi) =  0
\]
and
\[
 Q_{\bar\nu_p}^{\bar v/2}(0) \, P_{\bar\nu_p}^{\bar v/2} (\tanh k\pi) =  0 \ ,
\]
and, once again, no mode exists for $\bar m_p = 0$ and $\tilde m_p =0$.
Since the 
$n \neq 0$ modes 
decouple from low energy physics, the first non vanishing modes are
$A_{4,5}^{(0,1)}(x_{\mu})$.

\subsubsection{Bulk masses}
\label{sec:bulk_mass}
At this stage, let us reexamine the consequences of a nonzero bulk
mass term $M$ (irrespective of whether it has been occasioned by an
explicit or spontaneous breaking of the symmetry). 
The mass parameter $M$ makes its presence felt in the context of the 
four-dimensional masses $m_{np}$ primarily through the quantities $u$ 
and $\bar v$ (see eqns.(\ref{gauge_soln}, \ref{adjoint_soln4} \& \ref{adjoint_soln5})). 
For the large $k$ case, even if $M$ is of the order of the natural cutoff, 
the quantities
$u, v$ do not differ substantially from their $M=0$ limit values of 
$3$ and $1$ respectively.
Naively, this would
lead one to presume that the existence of even a seemingly large $M$
would not alter the low-lying part of the KK-tower spectrum to any
significant degree. However, as has been shown in ~\cite{chang}, this
argument is fraught with danger, and much the same follows here. As
can be appreciated easily, with $M$ being the bulk mass term, the mass
of any given mode $m_{np} > M$. Indeed, in the absence of warping,
we would expect $m_{00} = M$ and $m^2_{np} = M^2 + n^2 / R_y^2 +
p^2/r_z^2$. With $M^{-1}$ being of the same order as the
compactification radii, the mass of the first excitation, in the
absence of warping, would be of similar order as that of the zero
mode. Thus, if warping is supposed to bring down the mass of the zero
mode, as perceived in four-dimensions, from $M$ to $M_Z$ ($Z$-boson
mass), it would also, typically, bring down the first excitation to
well below a TeV, resulting in severe phenomenological
contradictions. In other words, a bulk $M$, if present, would need to
be much smaller than the compactification scale, thereby bringing back
the hierarchy problem in a new guise. A similar argument holds for the 
large $c$ case too.

All of the analysis presented above carries through for a nonabelian
theory as well. The additional features are the gauge boson
self-interactions (a discussion of which we postpone until
Sec.\ref{sec:interactions}) and ghost fields. The latter, we consider
next.

\subsection{Ghosts}
    \label{subsec:ghots}
As is well-known, a covariant gauge fixing condition 
for a nonabelian theory always gives rise to ghost fields. Accounting for the 
spontaneous symmetry breaking, the gauge-fixing term can be written as
\begin{equation}
\barr{rcl}
{\cal L}_{gf} & = & \dis \frac{\sqrt{-g}}{2 \zeta} \, G^a \, G^a 
\\[2ex]
G^a(A) & =  & \dis \left(
        \frac{\eta^{\mu\nu}}{a^2b^2}\partial_{\mu}A^{a}_{\nu} 
      + \left\{ \frac{\partial_{4}\left(a^2b A^{a}_4 \right)}{R_y^2a^2b^3} 
              + \frac{\partial_{5}\left(a^2b^3 A^{a}_5 \right)}{r_z^2a^2b^3} 
        \right\} 
   \right) -\zeta \gym T^{a}_{ij}\phi_{0i}\chi_{j},
\earr
\end{equation}
with $T^a_{ij}$ being the Yang Mills group generators in the 
representation of the scalar fields $\phi_i$. The latter are decomposed
into the vacuum values $\phi_{0i}$ and the perturbations $\chi_i$
  around it ($\phi_i = \phi_{0i} + \chi_i$). 
Evidently, the gauge boson mass matrix is given by 
\[
(M^{2})^{ab}  = \gym^2 T^{a}_{ij}T^{b}_{kj}\phi_{0i}\phi_{0k} \ .
\]
Note that the scalars $\phi_i$ need not be bulk fields, but may be 
confined to branes (as we shall argue for in Sec.\ref{sec:higgs}), with 
the appropriate delta-functions being included accordingly.  

Under a gauge transformation, 
\[
\barr{rclcl}
A^a_{M}& \rightarrow & 
A^{a\lambda}_{M} & = & \dis
A^{a}_{M}+\frac{1}{\gym}\partial_{M}\theta^{a}+f^{abc}A^{b}_{M}\lambda^{c} 
\\[1ex]
\phi_i & \rightarrow & \phi_i + \delta \phi_i & = & 
    \dis \phi_i - \lambda^{a}T^{a}_{ij}\phi_j
\earr
\]
where $f^{abc}$ are the structure constants.
Using the notations of ref.\cite{peskin}, 
the corresponding Faddev-Popov (FP) determinant is
\[
{\rm det} \left(\frac{\delta G^a(A^{\theta})}{\delta\lambda^b}\right)
= {\rm det} 
  \left(\frac{1}{\gym} \, 
         \left[
     \frac{\eta^{\mu\nu}}{a^2b^2}\partial_\mu{\cal D}_{\nu}^{ab} 
   + \frac{1}{a^2b^3}\left\{ 
               \partial_4(a^2b \, {\cal D}_4^{ab}) 
             + \partial_5(a^2b^3 \, {\cal D}^{ab}_5)
           \right\} 
   - (M^2)^{ab} + {\cal O}(\chi)\right]
\right)
\ ,  
\]
where ${\cal D}_{M}^{ac} = \partial_{M}\delta^{ac} + g f^{abc}
A^{b}_{M}$. Though they are important in their own right, 
since
a study of the higgs interactions does not constitute 
the main aim of this paper, we
neglect the $O(\chi)$ term.  Writing the determinant as a gaussian
integral over an anticommuting scalar $(\theta)$ in the adjoint
representation of the gauge group yields the Lagrangian for the ghost
field, namely
\[
{\cal L}_{g.f.} \, = \,
        \sqrt{-g} \, \bar{\theta}^{a}(x^{M}) \, 
               \left[ \Box^{ab}- (M^2)^{ab} \right] \, \theta^{b}(x^{M}) \ .
\]
The ghost kinetic term, then, is
\[
{\cal L}_{\rm gh.kin} = R_yr_z \, 
\bar{\theta}^{a}\left[a^2b^3 \eta^{\mu\nu}\partial_\mu \partial_\nu
   +\frac{a^4 \, b^3}{R_y^2} \left( \partial_4^{2}
                            +  \frac{2 \partial_4 a}{a} \partial_4 \right)
+\frac{a^4b^5}{r_z^2} \left( \partial_5^2 
    + \frac{3 \, \partial_5 b}{b} \, \partial_5\right)
     - (M^2)^{ab} \right] \theta^{b} \ .
\]
Decomposing, in anticipation, as
\beq
\theta^{a}(x_{\mu},x_4,x_5) = \frac{1}{\sqrt{R_y r_z}} \,
       \sum_{n,p} \theta^{a\, (n,p)}(x_{\mu}) \, \eta_{n,p}(x_4) \,
            \chi_p(x_5) \ ,
\eeq
it is easy to see that $\eta_{n,p}$ and and $\chi_p$ satisfy the same
equations of motion (and orthonormality properties) as the corresponding
wavefunctions for the vector modes (see eqns.(\ref{vector:ortho}\&
\ref{eqofmotionz})). In other words, not only are their masses identical 
to those for the corresponding vector modes, the wavefunction modes 
are the same too. This, of course, is as expected.

\section{Higgs}
   \label{sec:higgs}
In the preceding section, we had, for the sole purpose of determining
the spectrum, included an explicit mass term for the gauge boson,
without ascribing any dynamic origin to it. The simplest mechanism to
generate masses while preserving a gauge symmetry, of course, is to
introduce a Higgs field and effect a spontaneous breaking of the
symmetry.  The most straightforward, and seemingly natural, way to do
so would be to consider a bulk Higgs field. However, this immediately
leads to a problem (also seen in the context of bulk SM fields in the
RS scenario~\cite{chang}) in that the consequent masses of the 
first excited modes of the gauge
bosons typically turn out to be well below\footnote{The argument is 
exactly the same as the one in the preceding section arguing against a
naturally large bulk mass term (howsoever generated) for the gauge bosons.}
$1 \tev$, in stark contradiction to the 
direct bounds from gauge boson (most typically, generic $Z'$) 
searches\footnote{Some of these constraints on bulk Higgs 
can be evaded, though, {\em e.g.} if one considers soft-wall 
scenarios~\cite{soft_wall}. However, most such constructions, typically,
need additional inputs, whether it be in the form of  
gauge-Higgs unification or the localization of the Higgs close to the 
singularity, and, for certain models, even additional dynamics
to generate and stabilize the soft-wall setup itself.}.

The alternative, then, is to localize the Higgs on one of the
4-branes, or, perhaps even to a 3-brane at one of the junctions.  A
junction localized Higgs would lead to a gauge field equation of
motion analogous to a two dimensional Poisson equation with a point
source term. With the rotational symmetry being absent, a solution of
such an equation consistent with the boundary conditions, though, is a
complicated one. 

\subsection{3-brane localization}
Notwithstanding the anticipated technical problems,
we begin by considering 
a 3-brane localized Higgs with a 
Lagrangian given by
\beq
\barr{rcl}
{\cal L}_h & = & \dis \delta(x_4-\pi)\delta(x_5) \sqrt{-g_4}
    \Big( g^{\nu \rho}D_\nu \Phi(x^{\mu})^{\dagger} D_\rho \Phi(x^{\mu}) 
+ V(\Phi(x^{\mu})) \Big) \ ,
\\[2ex]
V(\Phi) & = & \dis \lambda \, \Big(\Phi^{\dagger} \Phi- \tilde v^2 \Big)^2 
\\[2ex]
D_{\mu} & = & \dis \partial_{\mu} -i \, \gym\, A_{\mu}(x_{\nu},x_4,x_5)
\earr
    \label{3brane_h_lagr}
\eeq
Rewriting $\Phi(x^{\mu}) = \left[\tilde v  + h(x^{\mu}\right] / \left[ a(\pi)b(0) \right]$
allows us to canonically quantize the Higgs field, with the 
Higgs mass being given by
\beq 
m_h = \sqrt{\lambda} \, \tilde v \, \frac{e^{-c\pi}}{\cosh(k
  \pi)} \equiv \sqrt{\lambda} \, v \ .  
\eeq
Instead of solving the
consequent equations of motion for the gauge fields with the delta
function sources included, we consider the latter to be localized
perturbations to the system with the symmetry intact.  This is a
  valid approximation as the effect of the spontaneous breaking of the
  gauge symmetry is parametrized by $v$; since it is suppressed down to
  the electroweak scale, its contribution to the KK-gauge boson masses
  would be small compared to those due to compactification (even after
  the warping).  To the first order, then, the mass spectra that we
computed in section(~\ref{mass spectra}) do not get disturbed in the
bulk, but for a small correction\footnote{This, of course, pertains
  only to the KK-excitations. For the zero-modes, this would be the
  only mass term.}  due to symmetry breaking in the brane. Similarly,
the wavefunction profile would change only close to the brane. 
With the new contribution to the gauge boson mass term being
\[
\int dx_4 dx_5 \gym^2\sqrt{-g} g^{\mu \nu}\delta(x_4-\pi)\delta(x_5)\Phi^{\dagger}\Phi A_{\mu}A_{\nu} \ ,
\]
it could be expressed in terms of the component fields (see eqn.\ref{gauge_boson_comp}) as 
\beq
m^2_{s.b.}(n_1, p_1; n_2,p_2) = \frac{\gym^2 \, v^2}{R_yr_z} \, 
           \eta_{n_1,p_1}(\pi) \,
             \eta_{n_2,p_2}(\pi) \, \chi_{p_1}(0) \, \chi_{p_2}(0) \ .
\label{SSB_gauge_mass}
\eeq
It comes as no surprise that this term mixes the KK modes, for this is  
a generic feature of brane localized Higgs fields. 
To the zeroth order, the gauge wavefunctions remain unchanged and have
no discontinuities on the branes. Of course, to obtain the 
physical gauge boson states and their masses, one would need to diagonalize 
the mass matrix including the terms of eqn.(\ref{SSB_gauge_mass}). While a 
closed form solution is not obvious, inspiration may be taken from the 
see-saw mechanism; 
although the hierarchy between the compactification 
and the Higgs contributions is not as large as in the neutrino sector, it is still sufficiently 
large for such approximate solutions to be valid. 
And while the absolute contribution of the 
Higgs grows 
for the higher-$p$ modes, primarily because $\chi_p$ are 
localized near $x_5 = 0$, 
this growth is relatively slow and consequently 
the ratio $m_{s.b.}/m_{np}$ becomes smaller. Thus,
neglecting the Higgs contribution for the higher modes is progressively a 
better approximation, and obtaining good estimates of the gauge boson 
masses and eigenstates is a relatively straightforward task.

Going beyond the first order approximation renders the 
algebra to be rather cumbersome, without providing us any 
real insight. To gain the latter, we would need to take recourse 
to some approximation. One such could be a smearing of the delta-function 
localized Higgs field.

\subsection{4-brane localization}
Choosing to work in the regime wherein the warping in the
$x_5$-direction is small, we consider a configuration such that the
factor $\delta(x_5)$ in eqn.(\ref{3brane_h_lagr}) is smeared onto a
nonsingular but localized function. This, of course, is equivalent to
considering the Higgs-field to be localized onto the 4-brane at $x_4 =
\pi$ (rather than on the 3-brane at the junction), with a further
concentration of its wavefunction close to $x_5 = 0$. The relevant
part of the symmetry breaking Lagrangian can now be parametrized as
\beq
{\cal L}_m = \sqrt{-g_5} \, 
         \frac{\widetilde M^2(x_5)}{2}g^{\mu \nu} A_{\mu}A_{\nu} \delta(y-\pi) 
\ ,
\eeq
with $\widetilde M(x_5)$ encapsulating the 
$x_5$-dependent profile of the Higgs vacuum expectation value.
Assuming, for the moment, that $\widetilde M = m / \sqrt{b(x_5)}$ where $m$ is a constant 
(we shall comment later on the origin of such a profile), 
the equations of motion for the gauge boson modes are altered from 
those in eqn.(\ref{eqofmotionz}) to
\begin{equation}
\label{eqofmotion1y}
\barr{rcl}
m^2 a^2 \delta(y-\pi) & = & \dis 
m_{np}^2  - m_p^2 a^2 + \frac{1}{R_y^2 \eta_{n,p}} \, 
              \partial_4 \big(a^2\partial_4 \eta_{n,p}  \big) 
\\[2ex]
-m_p^2 & = & \dis 
\frac{1}{r_z^2 b \chi_p} \, \partial_5 \big( b^3 \partial_5 \chi_p \big) \ .
\earr
\end{equation}
Indeed, the particular form of $\widetilde M(x_5)$ was 
chosen so as to render the equation of motion separable and, hence,
easily solvable.
The form of the second equation above is evidently the same as that in 
eqn.(\ref{eqofmotionz}) but for the bulk mass term $M$ in the latter.
As for the first equation, 
for a given $m_{np}$, it differs from its predecessor (see eqn.(\ref{eqofmotionz}))
only as far as the boundary term is concerned. Consequently, the bulk 
solutions 
are exactly the same as in eqn.(\ref{gauge_soln}) but with $u = 3$.
The boundary conditions on $\chi_p(x_5)$ remain exactly the same
as before (see Sec.\ref{sec:boundary_gauge_gen}), namely
\[
\chi_p'|_{x_5=0} = 0 = \chi_p'|_{x_5=\pi} \ ,
\]
and, thus, the $m_p$-spectrum is unchanged. To obtain the full spectrum,
we must consider the boundary conditions on $\eta_{n,p}$, and one of these now changes to accommodate the brane localized term, to wit,
\begin{equation}
\label{bcy1}
\eta'_{n,p}|_{x_4=0} = 0 \ , 
\quad {\rm and} \quad
\eta'_{n,p}|_{x_4=\pi} =  m^2 \, R_y^2 \, \eta_{n,p}(\pi) \ .
\end{equation}
Since $m_{p=0} = 0 $, we have, for the modes $\eta_{n0}$,
\[
J_0(e^{-c\pi}\alpha_{n0})\left[ 2 c \alpha_{n0} Y_0(\alpha_{n0}) 
           + R_y^2m^2Y_1(\alpha_{n0})\right]
= 
Y_0(e^{-c\pi}\alpha_{n0}) \left[ 2 c \alpha_{n0} J_0(\alpha_{n0}) 
+ R_y^2m^2J_1(\alpha_{n0})\right] \ ,
\] 
where, as before, $\alpha_{n0} \equiv m_{n0} \, R_y e^{c\pi} / c$.  Since the lightest mass mode is to be identified with the
$W/Z$ bosons, we have $\alpha_{00}  \ll 1$ (as $c
\sim 10$). Expanding the Bessel functions, we obtain
\[
 m^{2}_{00} \approx \frac{1}{2 \pi} m^2 \, e^{-2c \pi} \ .
\]
Clearly, for the $W$ boson, $ m^2 = g_2^2 v^2$, whereas 
for the $Z$ boson, $m^2 = (g_2^2 + g_1^2) v^2$,  with $g_{2,1}$ being
the weak and hypercharge coupling constants respectively. 

In essence, the new boundary condition on $\eta_{n,p}$ manifests 
itself in a change in the gauge boson spectrum through
\begin{equation}
\alpha_{np}J_{\nu_p-1/2}(\alpha_{np}) +
\left(\frac{1}{2} + \frac{R_y^2m^2}{2 c} \, - \nu_p\right) \, 
J_{\nu_p+1/2}(\alpha_{np})=0 \ ,
\end{equation}
as distinct from eqn.(\ref{gauge_KK}).
With there being no singularities in 
the $x_5$--direction, all the $\chi_p$ modes remain unchanged (in particular, 
$\chi_0$ is flat), and all effects of the brane localized mass term 
for the gauge bosons manifest themselves in altering the quantized 
values of $\alpha_{np}$, and, hence, in the form of
$\eta_{n,p}$. Quite in parallel 
to the RS case~\cite{csabaeffective}, the change 
in the wavefunction for the zero modes ($W^{\pm (0,0)}, Z^{(0,0)}$) 
is concentrated close to the brane and, in magnitude, restricted to ${\cal O}(M_W^2 )$. Away from the brane, even this relative 
change falls off exponentially.

As can be gleaned from the discussions in this
 section, 
a generic brane localized Higgs profile would lead to 
equations for the gauge bosons that do not admit simple 
closed form solutions. While the particular choice of 
$\langle \phi \rangle = v / \sqrt{b(x_5)}$ may seem an ad hoc
one, we end this section delineating a mechanism to achieve this.
 The effective potential for a 
scalar field $\phi$ localized on the 
4-brane at $x_4 \, = \, \pi$
could be written as
\[
V_{\rm eff} = - r_z \, a^4(\pi) \, 
b^4 \, \left[ \frac{1}{2 r_z^2}\partial_5\phi \partial_5 \phi + V(\phi) \right] \ ,
\]
where $V(\phi)$ is the potential term appearing in the (flat space) Lagrangian.
To find what potential would give rise to the required form for
$\langle \phi \rangle$, we may treat 
$V_{\rm eff}$ as the Lagrangian for a 
one-dimensional particle. Varying with respect to $\phi$, we get
\[
\frac{1}{r_z^2}\partial_5(b^4\partial_5 \phi)  - b^4 V'(\phi) =0 \ .
\]
Demanding that the solution be of the form $\phi \propto 1 / \sqrt{b(x_5)}$, we have
\[
V(\phi) = \frac{k^2}{r_z^2}\Big( \frac{5 \; {\rm sech}^2 k \pi}{24 \,v^4} \,
      \phi^6 - \, \frac{7}{8} \, \phi^2 \Big) \ .
\]
At first sight, it might seem disquieting that the form of
  $V(\phi)$ is fixed so uniquely. However, as previous experience has
  shown, such is often the case if exact closed form solutions to a
  complicated system (such as the gravity-scalar system under
  consideration here). Deviations from the form above are, of course,
  permissible, but only at the cost of increasing complexity of the
  solution (or numerical approximations). We desist from exploring
  such possibilities as these do not make qualitative differences to
  the main features of interest.

\section{ Interactions}
\label{sec:interactions}
\subsection{Gauge--fermion}
The relevant piece of the
Dirac Lagrangian for the positive chirality six-dimensional 
field is given by
\[
{\cal L} \ni i \, \gym \, \sqrt{-g} \, \bar \Psi_{+} E^{\mu}_a 
                                      \Gamma^{a} \Psi_+ A_{\mu} \ ,
\]
where the group representation has been suppressed. (A similar 
account holds for the negative chirality field as well.)
Writing the term above 
in its component form, we have
\beq
{\cal L} \ni \sum_{\{n_i,p_i\}}
        g^{V,f}_{\{n_i, p_i\}} \bar \psi^{n_1,p_1}_{l/r}\gamma^{\mu} 
                \psi^{n_2,p_2}_{l/r}  A^{n_3,p_3}_{\mu} \ ,
\eeq
with the four dimensional charges being given by
\begin{equation} 
\label{charge}
\dis g^{V,f}_{\{n_i, p_i\}} 
= \frac{\gym}{\sqrt{R_yr_z}}\int_0^\pi d x_4 \, \int_{-\pi}^{\pi} d x_5 \,
      a^3b^4
    {\cal F}^{n_1,p_1}_{l/r}(x_4,x_5) \, {\cal F}^{n_2,p_2}_{l/r}(x_4,x_5)
        \, \eta_{n_3,p_3}(x_4) \, \chi_{p_3}(x_5) \ .
\end{equation}
It is evident that, for the gauge boson zero mode, we have
\beq
 \dis
   g^{V,f}_{\{n_i, p_i\}} \to 
\frac{\gym}{\sqrt{R_y r_z}}\sqrt{\frac{k}{2 \pi \tanh(k \pi)}} \delta_{n_1,n_2} \delta_{p_1,p_2}\ .
    \label{zero_mode_coup}
\eeq
This universal coupling of the zero mode is, of course, 
mandated by gauge invariance. 

For the adjoint scalar $A_4$, we start from
\[
{\cal L} \ni i \gym \, \sqrt{-g} \, \bar \Psi_{+}E^{4}_a \Gamma^{a} 
                  \Psi_+ A_{4} \ .
\]
Rewriting in terms of the four-dimensional fields, we have
\beq
{\cal L} \ni \sum_{\{n_i, p_i\}} g_{4,f\{n_i, p_i\}} 
             \bar \psi^{n_1,p_1}_{l/r}\gamma^{5} \psi^{n_2,p_2}_{r/l} \, 
           \tilde{A}^{n_3,p_3}_{4} \ ,
\eeq
where the four dimensional coupling constants are given by
\begin{equation}
\label{chargeadjointY}
g_{4,f\{n_i, p_i\}}  =   \frac{\gym}{\sqrt{R_yr_z}} \,
      \int_0^\pi d x_4 \, \int_{-\pi}^{\pi} d x_5 \,
            a^4b^4 \, {\cal F}^{n_1,p_1}_{l}(x_4,x_5) \,
                     {\cal F}^{n_2,p_2}_{r}(x_4,x_5) \,
                        \eta^{(4)}_{ n_3,p_3}(x_4) \, \chi^{(4)}_{p_3}(x_5) \ .
\end{equation}

Similarly, for the adjoint scalar $A_5$, we start from
\[
{\cal L} \ni i \gym \, \sqrt{-g} \, \bar \Psi_{+}E^{5}_a \Gamma^{a} 
                  \Psi_+ A_{5} \ .
\]
Redefining (as before) $\frac{R_y}{r_z}A_5 \rightarrow \tilde{A_5}$, and 
writing in terms of the components, we have 
\beq
{\cal L} \ni \sum_{\{n_i, p_i\}} g_{5,f\{n_i, p_i\}} 
             \bar \psi^{n_1,p_1}_{l/r} \psi^{n_2,p_2}_{r/l} \, 
           \tilde{A}^{n_3,p_3}_{5} \ ,
\eeq
where the four dimensional coupling constants are given by
\begin{equation}
\label{chargeadjointZ}
g_{5,f\{n_i, p_i\}}  =   \frac{\gym}{\sqrt{R_yr_z}} \,
      \int_0^\pi d x_4 \, \int_{-\pi}^{\pi} d x_5 \,
            a^4b^5 \, {\cal F}^{n_1,p_1}_{l}(x_4,x_5) \,
                     {\cal F}^{n_2,p_2}_{r}(x_4,x_5) \,
                        \eta^{(5)}_{n_3,p_3}(x_4) \, \chi^{(5)}_{p_3}(x_5) \ .
\end{equation}

\begin{table}[!h]
\begin{tabular}{ccc}
$
\begin{array}{|c|c|c|}
\multicolumn{3}{c}{\underline{k = 0.3, \; \alpha  = 49, \;  w = 1.82 \times 10^{-14}}}
\\[1ex]
\hline
(n,p) & m_{np} (\tev) & C_{np}  \\
\hline 
(1,0)&  4.47   & 3.87  \times 10^{0} \\ 
\hline 
(2,0) &  10.2  & 4.98   \times 10^{-1}   \\ 
\hline 
(0,1)& 10.1  & 7.89  \times 10^{-1}   \\ 
\hline 
(1,1) &  17.0 & 3.03   \times 10^{-1}  \\ 
\hline 
\end{array}
$
& \hspace*{3em} &
$
\begin{array}{|c|c|c|}
\multicolumn{3}{c}{\underline{k = 0.56, \; \alpha  = 50.4, \; w = 4.48 \times 10^{-14}}}
\\[1ex]
\hline
(n,p) & m_{np} (\tev) & C_{np} \\
\hline 
(1,0)&   8.55    & 3.77 \times 10^{0} \\ 
\hline 
(2,0) &  19.6  & 4.93 \times 10^{-1}   \\ 
\hline 
(0,1)&  14.6   & 2.35  \times 10^{0}  \\ 
\hline 
(1,1) &  26.9  &  7.19  \times 10^{-1} \\ 
\hline 
\end{array}
$
\end{tabular}
\caption{\em Sample spectra for the small $k$ case 
for a particular bulk curvature ($\epsilon = 0.0775$) with $R_y$ set to satisfy the hierarchy eqn.(~\ref{hierarchy}).
The ratio $C_{np}$ ( coupling of vector gauge boson to massless fermion bilinear)
 is as defined in eqn.(\ref{coup_scaling}).}
\label{tab:small_k}
\label{smallktable}
\end{table}

What is of particular interest, especially in the context of
  collider searches, is the coupling of a relatively low-lying KK gauge
  boson to a pair of SM fermions (in other words, the zero
  modes). Some examples of gauge boson
spectra and their couplings to the lowest modes of the
fermion current are given in Tables
\ref{smallktable} \& \ref{largektable}.  The measure of importance, 
apart from the mass of the level-$(n,p)$ KK gauge boson mass, is the scaling 
$C_{np}$ of its coupling with the SM fermions, viz.
\beq
   C_{np} \equiv \frac{g^{V,f}_{\{0,0,n\},\{0,0,p\}}}
                     {g^{V,f}_{\{0,0,0\},\{0,0,0\}}} \ .
\label{coup_scaling}
\eeq

\begin{figure}[!h]
\vspace*{-50pt}
\centerline{\epsfxsize=8cm\epsfbox{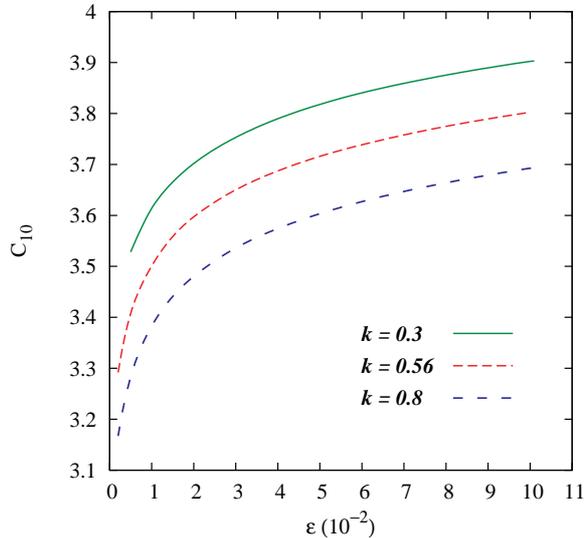}  }
\vspace*{-70pt}
\caption{\em 
The matter coupling $C_{10}$
to the first vector gauge mode as a function of $\epsilon$ for a fixed $k$.}
\label{fig:small_k_mass&coup}
\end{figure}

Concentrating on the small $k$ scenario (Table
  \ref{smallktable}), it is interesting to note the relative closeness
  of the excitations (beyond the first one) as compared to the RS case
  with bulk gauge bosons and fermions.  And as in the latter case (and
  unlike in the flat extra-dimensional scenarios), the coupling of a SM fermion pair
  to the gauge excitations are not universal. In
  particular, the coupling to the first excitation is enhanced
  compared to that for the zero mode, while those to the higher ones
  are suppressed. Furthermore, the enhancement for the $(1,0)$ mode is
  only slightly smaller than that for the corresponding
  five-dimensional theory, with this effect having only a marginal
  dependence on the value of $k$.  This is not surprising since the
  leading dependence of the couplings on $k$ is common to the ground
  and the $(1,0)$ state. On the other hand, if the extent of warping
  is kept constant, the value of $c$ progressively decreases from the
  RS value as $k$ increases from zero. Consequently, $C_{np}$
  decreases, although only slowly. This is shown in 
  Fig.~\ref{fig:small_k_mass&coup}.
  Although a substantial decrease in
  $C_{10}$ is possible (thereby making these bosons less accessible
  to collider searches etc.), that would require a
  relatively large $k$. However, as already mentioned, if $c$ and $k$
  are to be of the same order, a very large hierarchy between $R_y$
  and $r_z$ would be required, thereby bringing back the hierarchy
  problem in a different guise. In other words, the aesthetically 
pleasant region of the parameter space would lead to gauge bosons discoverable 
in the next run of the LHC.

The situation is very different in the large $k$ regime. Since the
warping in the $x_4$-direction is very small, the $n \neq 0$ modes are
all super-heavy and decouple from the TeV scale physics. In other
words, just like the situation for the gravitons~\cite{Arun:2014dga},
essentially only one tower of gauge bosons (or, fermions for that
matter) remains. Moreover, the couplings of the KK-tower are enhanced
to nonperturbative levels. This is demonstrated, for a moderate hierarchy,
in
Table.\ref{largektable}. Pushing the
fundamental scale $M_6$ even higher would, typically, result in the
couplings growing even further, with the only way out of this
eventuality being the introduction of a big hierarchy between the
moduli. Indeed, one had seen the germ of this problem even in the
graviton sector~\cite{Arun:2014dga}.  Thus, bulk fermions or gauge
bosons in the large $k$ regime is ill-suited for a perturbative
treatment.

\begin{table}[!h]
\begin{center}
\begin{tabular}{ccc}
$
\begin{array}{|c|c|r|}
\multicolumn{3}{c}{\underline{k=6, \; \alpha = 16.8, \; \epsilon = 0.001}}
\\[1ex]
\hline
(n,p) & m_{np} (\tev) & C_{np}  \\ 
\hline 
(0,1) &  22.0   & 1.71 \times 10^{3}    \\ 
\hline 
(0,2) & 49.1    & 3.06 \times 10^{2}  \\ 
\hline  
(0,3) &  75.0  &  1.29 \times 10^{2}\\ 
\hline 
\end{array}
$
\end{tabular}
\end{center}
\caption{\em Sample spectrum for the large $k$ case with $R_y$ set to satisfy the hierarchy eqn.(~\ref{hierarchy}). The ratio $C_{np}$ 
 is as defined in eqn.(\ref{coup_scaling}).}
\label{largektable}
\end{table}


\subsection{Triple gauge boson couplings}
The trilinear self interaction term 
can be written as 
\[
\barr{rcl}
{\cal L} & \ni & \dis \gym f^{abc} \left[ 
   R_y r_z \, b \eta^{\mu\alpha}\eta^{\nu \rho}
     (\partial_\mu A^{a}_\nu) A^{b}_\rho A^{c}_\alpha 
   + \frac{r_z}{R_y} a^2b \eta^{\mu \nu}
      \left\{ (\partial_\mu A^{a}_4) A^{b}_4 A^{c}_\nu 
            + (\partial_4 A^{a}_\mu)A^{b}_4 A^{c}_\nu \right\}
     \right.
\\[1ex]
   & & \dis \hspace*{6em} \left.
   + \frac{R_y}{r_z} a^2b^3 \eta^{\mu \nu}
      \left\{ (\partial_\mu A^{a}_5) A^{b}_5 A^{c}_\nu 
            + (\partial_5 A^{a}_\mu)A^{b}_5 A^{c}_\nu \right\}
\right]  \ ,
\earr
\]
where $f^{abc}$ are the structure constants. The six-dimensional Ward
identities ensure that the coupling $\gym$ equals the 
six-dimensional gauge-fermion coupling even after quantum corrections 
are included. Rewriting in terms of the four dimensional fields, we have
\beq
\barr{rcl}
\dis {\cal L} \ni f^{abc} \, \sum_{\{n_i, p_i\}} & \Big[  
& \dis 
g^{(3v)}_{\{n_i, p_i\}} \eta^{\mu\alpha}\eta^{\nu \rho}
     \left(\partial_\mu A^{a(n_1,p_1)}_\nu\right) A^{b(n_2,p_2)}_\rho 
                  A^{c(n_3,p_3)}_\alpha 
\\[0ex]
& + & \dis
g^{(vv4)}_{\{n_i, p_i\}}   \eta^{\mu \nu} 
      A^{a(n_1,p_1)}_\mu A^{b(n_2,p_2)}_\nu \tilde{A}^{c(n_3,p_3)}_4 
\\[0ex]
& + & \dis
 g^{(44v)}_{\{n_i, p_i\}}   \eta^{\mu \nu} 
   \left(\partial_\mu \tilde{A}^{a(n_1,p_1)}_4\right) \tilde{A}^{b(n_2,p_2)}_4 
         A^{c(n_3,p_3)}_\nu 
\\[0ex]
& + & \dis
g^{(vv5)}_{\{n_i, p_i\}}   \eta^{\mu \nu} 
      A^{a(n_1,p_1)}_\mu A^{b(n_2,p_2)}_\nu \tilde{A}^{c(n_3,p_3)}_5 
\\[0ex]
& + & \dis
 g^{(55v)}_{\{n_i, p_i\}}   \eta^{\mu \nu} 
   \left(\partial_\mu \tilde{A}^{a(n_1,p_1)}_5\right) \tilde{A}^{b(n_2,p_2)}_5 
         A^{c(n_3,p_3)}_\nu 
\Big] \ ,
\earr
\eeq
where the coupling constants are defined through
\beq
\barr{rcl}
\dis 
g^{(3v)}_{\{n_i, p_i\}} & = & \dis
\frac{\gym}{\sqrt{R_yr_z}}
   \int_0^\pi d x_4 \, \eta_{n_1,p_1} \, \eta_{n_2,p_2} \, \eta_{n_3,p_3} \;
   \int_{-\pi}^{\pi} d x_5 \, b \chi_{p_1} \, \chi_{p_2} \, 
                \chi_{p_3} \ ,
\\[2ex]
\dis g^{(vv4)}_{\{n_i, p_i\}} & = & \dis
\frac{\gym }{R_y^2}\sqrt{\frac{R_y}{r_z}}
   \int_0^\pi d x_4 \, a^2 \eta_{n_1,p_1} \, \eta_{n_2,p_2} \, 
                                             \eta^{(4)}_{n_3, p_3}
   \int_{-\pi}^{\pi} d x_5 \, b (\partial_4\chi_{p_1}) \, 
                     \chi_{p_2} \, \chi^{(4)}_{p_3} \ ,
\\[2ex]
\dis g^{(44v)}_{\{n_i, p_i\}} & =  & \dis 
\frac{\gym}{\sqrt{R_y r_z}} \, 
  \int_0^\pi d x_4 \, a^2 \eta^{(4)}_{n_1,p_1} \, \eta^{(4)}_{n_2,p_2} \, 
                           \eta_{n_3,p_3} \;
  \int_{-\pi}^{\pi} d x_5 \, 
        b \chi^{(4)}_{p_1} \, \chi^{(4)}_{p_2} \, \chi_{p_3} \ ,
\\[2ex]
\dis g^{(vv5)}_{\{n_i, p_i\}} & = & \dis
\frac{\gym}{r_z^2}\sqrt{\frac{r_z}{R_y}}
   \int_0^\pi d x_4 \, a^2 \eta_{n_1,p_1} \, \eta_{n_2,p_2} \, 
                                             \eta^{(5)}_{n_3, p_3}
   \int_{-\pi}^{\pi} d x_5 \, b^3 (\partial_5\chi_{p_1}) \, 
                     \chi_{p_2} \, \chi^{(5)}_{p_3} \ ,
\\[2ex]
\dis g^{(55v)}_{\{n_i, p_i\}} & =  & \dis 
\frac{\gym}{\sqrt{R_y r_z}} 
  \int_0^\pi d x_4 \, a^2 \eta^{(5)}_{n_1,p_1} \, \eta^{(5)}_{n_2,p_2} \, 
                           \eta_{n_3,p_3} \;
  \int_{-\pi}^{\pi} d x_5 \, 
        b^3 \chi^{(5)}_{p_1} \, \chi^{(5)}_{p_2} \, \chi_{p_3} \ .
\earr
\eeq
For the three vector vertex, clearly if one of them
is a zero-mode, the other two must be identical. Similarly, the vector zero-mode
couples only to a pair of identical scalars.
Finally, for either case, the coupling is the same as that in eqn.(\ref{zero_mode_coup}). All of the above are, of course, consequences of gauge invariance.
Finally, although there exists $\tilde A_4$--$\tilde A_5$ mixing term, 
it, as discussed earlier, is rather subdominant, and we 
omit it here.

\subsection{Quartic gauge interaction}
The corresponding term in the Lagrangian is 
\[
{\cal L}\ni \gym^2 f^{abc}f^{ced}R_y r_z  \, \eta^{\mu \rho} \, 
             A^{a}_{\mu} \, A^{e}_{\rho} \, 
     \left( b \, \eta^{\nu \alpha} \, A^{b}_{\nu} \, A^{d}_{\alpha} 
         + \frac{a^2b}{R_y^2}A^{b}_{4} A^{d}_{4} +  \frac{a^2b^3}{r_z^2} A^{b}_{5} A^{d}_{5} \right) \ .
\]
Once again, reexpressing in terms of four 
dimensional fields, we have
\beq
\barr{l}
\dis {\cal L} \ni f^{abc}f^{ced} \,  \eta^{\mu \rho} \, 
      \sum_{\{n_i, p_i\}} A^{a(n_1,p_1)}_{\mu} \, A^{e(n_3,p_3)}_{\rho} \, 
         \left[ {\cal G}^{(4v)}_{\{n_i, p_i\}} \, 
                          \eta^{\nu \alpha} \, A^{b(n_2,p_2)}_{\nu} \,
                                               A^{d(n_4,p_4)}_{\alpha}
    \right.
\\[2ex]
\hspace*{10.5em} \left. 
+ {\cal G}^{4(2v,2s)}_{\{n_i, p_i\}} \, 
                       \tilde{A}^{b(n_2,p_2)}_{4} \, \tilde{A}^{d(n_4,p_4)}_{4} 
+ {\cal G}^{5(2v,2s)}_{\{n_i, p_i\}} \, 
                       \tilde{A}^{b(n_2,p_2)}_{5} \, \tilde{A}^{d(n_4,p_4)}_{5} 
\right] \,
\earr
\eeq
with the coupling constants being defined through
\beq
\barr{rcl}
\dis {\cal G}^{(4v)}_{\{n_i, p_i\}} & = & \dis
\frac{\gym^2}{R_y r_z}\,
   \int_0^\pi d x_4 \eta_{n_1,p_1} \, \eta_{n_2,p_2} \,
                    \eta_{n_3,p_3} \, \eta_{n_4,p_4} \;
   \int_{-\pi}^{\pi}d x_5 \, b \, \chi_{p_1} \, \chi_{p_2} 
                               \, \chi_{p_3} \, \chi_{p_4} \ ,
\\[2ex]
\dis {\cal G}^{4(2v,2s)}_{\{n_i, p_i\}} & = & \dis
\frac{\gym^2}{R_y r_z}\,
      \int_0^\pi d x_4 \, a^2 \, \eta_{n_1,p_1} \, \eta_{n_3,p_3}
                              \, \eta^{(4)}_{n_2,p_2} \, \eta^{(4)}_{n_4,p_4} \;
      \int_{-\pi}^{\pi} d x_5 \, b \,
                \chi_{p_1} \, \chi_{p_3} 
              \, \chi^{(4)}_{p_2} \, \chi^{(4)}_{p_4} \ ,
\\[2ex]
\dis {\cal G}^{5(2v,2s)}_{\{n_i, p_i\}} & = & \dis
\frac{\gym^2}{R_y r_z}\,
      \int_0^\pi d x_4 \, a^2 \, \eta_{n_1,p_1} \, \eta_{n_3,p_3}
                              \, \eta^{(5)}_{n_2,p_2} \, \eta^{(5)}_{n_4,p_4} \;
      \int_{-\pi}^{\pi} d x_5 \, b^3 \,
                \chi_{p_1} \, \chi_{p_3} 
              \, \chi^{(5)}_{p_2} \, \chi^{(5)}_{p_4} \ .
\earr
\eeq
Again, for the zero mode vectors, the Ward identity is satisfied.

\subsection{Ghost vertices}
The relevant piece in the ghost Lagrangian is
\[
{\cal L}_{gh} \ni \gym f^{abc} R_yr_z \,
    \bar{\theta}^{a} \,
   (a^2b^3\eta^{\mu\nu}\partial_{\mu}A^{b}_{\nu} 
        +\frac{1}{R_y^2} a^3b^3(\partial_4a) A_4^{b}
        +\frac{1}{R_y^2} a^4b^3 \partial_4A_4^{b}
 \]
 \[
        +\frac{1}{r_z^2}a^4b^5\partial_5A^{b}_5
        +\frac{3}{r_z^2}a^4b^4(\partial_5b)A^{b}_5)
 \theta^{c} \ .
\]
yielding in terms of the four dimensional fields 
\beq
\barr{rl}
\dis {\cal L}_{gh} \ni f^{abc} \sum_{\{n_i, p_i\}} \bar{\theta}^{a(n_1,p_1)} \,
         \theta^{c(n_2,p_2)} 
    & \left[ g^{(1)}_{\{n_i, p_i\}} \,
            \eta^{\mu\nu} \partial_{\mu} A^{b(n_3,p_3)}_{\nu} 
            + g^{(2)}_{\{n_i, p_i\}} \, 
             \tilde{A}^{b(n_3,p_3)}_4 
\right.
\\[0ex]
&  \left.        
            + g^{(3)}_{\{n_i, p_i\}} \, \tilde{A}^{b(n_3,p_3)}_4 
      + g^{(4)}_{\{n_i, p_i\}} \, 
             \tilde{A}^{b(n_3,p_3)}_5 
          + g^{(5)}_{\{n_i, p_i\}} \, \tilde{A}^{b(n_3,p_3)}_5 
\right] \ , 
\earr
\eeq
where the coupling constants are defined as
\beq
\barr{rcl}
\dis g^{(1)}_{\{n_i, p_i\}} & = & \dis
\frac{\gym}{\sqrt{R_yr_z}} \int_0^\pi d x_4 \, a^2 \, \eta_{n_1,p_1} \, \eta_{n_2,p_2} \,
                                  \eta_{n_3,p_3} \; 
       \int_{-\pi}^{\pi} d x_5 \, b^3 \, 
                  \chi_{p_1} \, \chi_{p_2} \, \chi_{p_3} \ ,
\\[2ex]
\dis g^{(2)}_{\{n_i, p_i\}} & = & \dis
\frac{\gym}{R_y\sqrt{R_yr_z}} \, 
    \int_0^\pi d x_4 \, a^3 a' \, \eta_{n_1,p_1} \, \eta_{n_2,p_2} \, 
                                \eta^{(4)}_{n_3,p_3} \;
    \int_{-\pi}^{\pi} d x_5 \, b^3 \,
              \chi_{p_1} \, \chi_{p_2} \, \chi^{(4)}_{p_3} \ ,
\\[2ex]
\dis g^{(3)}_{\{n_i, p_i\}} & = & \dis
\frac{\gym}{R_y\sqrt{R_yr_z}} \, 
    \int_0^\pi dx_4 \, a^4 \, 
            \eta_{n_1,p_1} \, \eta_{n_2,p_2} \, \partial_4\eta^{(4)}_{n_3,p_3} \;
    \int_{-\pi}^{\pi} d x_5 \; b^3 \,
             \chi_{p_1} \, \chi_{p_2} \, \chi^{(4)}_{p_3} \ ,
                  
\\[2ex]
\dis g^{(4)}_{\{n_i, p_i\}} & = & \dis
\frac{\gym}{r_z\sqrt{R_yr_z}} \, 
    \int_0^\pi d x_4 \, a^4 \, \eta_{n_1,p_1} \, \eta_{n_2,p_2} \, 
                                \eta^{(5)}_{n_3,p_3} \;
    \int_{-\pi}^{\pi} d x_5 \, b^5 \,
              \chi_{p_1} \, \chi_{p_2} \, \partial_5\chi^{(5)}_{p_3}
    \ ,
\\[2ex]
\dis g^{(5)}_{\{n_i, p_i\}} & = & \dis
\frac{\gym}{r_z\sqrt{R_yr_z}} \, 
    \int_0^\pi dx_4 \, a^4 \, 
            \eta_{n_1,p_1} \, \eta_{n_2,p_2} \, \eta^{(5)}_{n_3,p_3} \;
    \int_{-\pi}^{\pi} d x_5 \; b^4 \,
            \dot b \,  \chi_{p_1} \, \chi_{p_2} \, \chi^{(5)}_{p_3} \ .
\earr
\eeq
Once, again, $g^{(1)}_{\vec 0, \vec 0} = g^{(3v)}_{\vec 0, \vec 0}$, 
as is mandated by gauge invariance.

\section{Summary}
\label{sec:summary}

While the negative results for graviton resonance searches by the
ATLAS~\cite{Aad:2012cy} and CMS~\cite{Khachatryan:2014gha}
collaborations have caused a bit of tension for the
Randall-Sundrum scenario, a six-dimensional analogue with a nested double
warping~\cite{Choudhury:2006nj} has been seen to be very consistent
with the experimental results~\cite{Arun:2014dga}. Such scenarios are
of interest in their own right as they could, for example, constitute an
intermediate step in the compactification down from a theory in higher
dimensions.  Moreover, the fundamental scale in such theories are
naturally lower than the Planck-scale, and this could play a
significant role in the context of gauge unification. But, most 
interestingly, it provides a tunable parameter that smoothly takes one from 
a nearly-conformal theory to another that is a large departure from 
one, with the added feature that both the ends provide a resolution 
of the hierarchy problem (although this is not apparent in the interim 
regime).

Just as the RS model would, generically, admit operators that lead to
unsuppressed flavour changing neutral currents and/or proton decay, so
would the model considered in
Refs.~\cite{Choudhury:2006nj,Arun:2014dga}. On a different vein, the
exact cutoff scale of this theory (normally described as the scale at
which the loop contributions are to be cutoff) needs to be identified
too.  It has been argued that, within the five-dimensional context,
the addition of the Planck-brane and/or the TeV-brane allows a
holographic interpretation~\cite{malda}, with the former acting as a
regulator leading to an ultraviolet\footnote{An analogous
argument for our case would imply a cutoff $\Lambda_{UV} \simeq
\mbox{min}(R_y^{-1}, r_z^{-1})$ as argued in Ref.~\cite{Arun:2014dga}
from an entirely different perspective.}  cutoff ($\lapp r_c^{-1}$)
on the corresponding
CFT~\cite{ArkaniHamed:2000ds,Rattazzi:2000hs,PerezVictoria:2001pa}.
It has been demonstrated that, for RS-like theories with gauge fields
extended in to the warped bulk, this is indeed
so~\cite{Pomarol:1999ad,Rizzo,Agashe:2002jx}.
Even though no such duality has been constructed for the case under consideration, it is quite conceivable that one such would exist. In the large 
$k$ case, the bulk is indeed $AdS_6$--like. However, for the 
phenomenologically more interesting case of large $c$ (small $k$),
 it is evident that the the metric is not conformally flat and, hence, 
a holographic interpretation would be considerably more tricky.

An alternative and obvious way to ameliorate flavour changing currents 
is to allow the fermions
(and, hence, the gauge fields too) to
propagate into the six dimensional bulk, for now the higher
dimensional operators get suppressed by a factor $\Lambda_{\rm UV}^4$, 
with  $\Lambda_{\rm UV}$ being the bulk cutoff of the theory,
which is higher than the GUT scale. Furthermore, a six-dimensional
theory allows one to make predictions about the number of chiral
generations in the theory.

An immediate consequence of taking these fields into the 
full six-dimensional bulk is that each of the KK-towers that are 
so familiar in the five-dimensional context now expand into a
``tower of towers'', thereby enriching the phenomenology, whether 
it be in the context of quantum correcions to SM amplitudes 
or direct production at, say the next run of the LHC.
In this paper, we have derived the wave profiles for these fields and
computed the master formula to calculate their spectra. It is seen
that, of the two branches allowed to the theory by the resolution of
the hierarchy problem, the one close to a conformally flat space leads
to the collapse of the `tower of towers' (for both the gauge fields
and the fermions) to a single tower each (the other excitations are
too heavy and decouple from the low-energy theory).  However, 
the spacing between the successive members of a tower is distinctly 
different from that in the five-dimensional analogue, thereby distinguishing
between the two scenarios. More tellingly though, the higher
KK-excitations of the gauge bosons couple very strongly,
thereby invalidating a perturbative treatment, and calls for a more 
sophisticated approach.

The other branch of the theory, which
cannot be thought of as a mere perturbation of the AdS$_6$ 
paradigm, retains the full tower of towers, with some of the excitations 
possibly being rather closely spaced. This naturally leads to
very interesting phenomenology, not the least of which 
pertains to the just-begun run of the LHC. It also  
is  of paramount interest in the
context of electroweak precision tests, rare decays and
renormalization group evolutions~\cite{inprep}. A particularly interesting feature is that the first KK mode for the fermions as well as the adjoint 
scalar is much heavier than that for the gauge boson. This is very different
from the case of the five-dimensional theory and constitutes a remarkable 
discriminant between the theories, say at the LHC.

As for the Higgs field, just as in the five-dimensional theory,
putting the Higgs field along with symmetry breaking potential
into the bulk either brings back the hierarchy problem, or renders the
masses of the gauge boson KK-tower unacceptably low. Thus, it is wise
to localize the Higgs on to a 3-brane.  This, though, has the
unfortunate consequence of making the combination of the equations of
motion and the boundary condition too complicated to permit an easy
understanding of the dynamics. On the other hand, if we localize the
Higgs onto a 4-brane (located at $x_4\, = \, \pi$ and $x_5\, = \, 0$), and
work in the large $c$ regime, we see that the Higgs vacuum expectation
value does get warped down to the electroweak scale.  With the
consequent brane localized contribution being small, it can be treated
as a perturbation, and the consequent shifts in the spectrum as well
as the wavefunction profiles can be calculated.  It is interesting to
note that, although small, the brane localized Higgs mixes the KK
states with possible phenomenological ramifications. These issues are
under investigation.

\section*{Acknowledgement}
   MTA would like to thank UGC-CSIR, India for assistance under Senior Research Fellowship Grant   
     Sch/SRF/AA/139/F-123/2011-12. DC acknowledges partial support from the  European Union FP7  ITN INVISIBLES (Marie Curie Actions, PITN- GA-2011- 289442), and the 
     Research and Development grant of the University of Delhi.

\end{document}